\RequirePackage[l2tabu, orthodox]{nag}
\RequirePackage{snapshot}

\documentclass[10pt,onecolumn]{extarticle}

\makeatletter
\if@twocolumn
  \usepackage[dvips,letterpaper,top=0.5in, bottom=0.5in, left=0.75in, right=0.5in,includefoot,heightrounded]{geometry}
\else
  \usepackage[dvips,letterpaper,margin=1in,includefoot,heightrounded]{geometry}
\fi

\usepackage{srcltx}

\usepackage[russian,portuges,english]{babel}

\iflanguage{portuges}
    {\newcommand{\keywordname}{Palavras-chaves}}
    {\newcommand{\keywordname}{Keywords}}

\usepackage{amsmath}
\usepackage{amssymb,amsfonts}

\usepackage{abstract}

\usepackage{graphicx}
\usepackage[usenames,dvipsnames,svgnames,x11names]{xcolor}
\usepackage{subfigure}

\usepackage{booktabs}

\usepackage{setspace}
\usepackage{flushend}

\usepackage{cite}

\usepackage{hyperref}
\usepackage[normalem]{ulem}
\usepackage{bm}

\usepackage{enumerate}

\usepackage{multirow}
\usepackage[noend]{algpseudocode}

\usepackage{listings}

\usepackage[short,12hr]{datetime}
       \usepackage{fouriernc}
\makeatletter

\newcommand{\printtitle}{%
\makeatletter
\if@twocolumn

\twocolumn[%
  \maketitle
  \begin{onecolabstract}
    \myabstract
  \end{onecolabstract}
  \begin{center}
    \small
    \textbf{\keywordname}
    \\\medskip
    \mykeywords
  \end{center}
  \bigskip
]
\saythanks
\else
  \maketitle
  \begin{onecolabstract}
    \myabstract
  \end{onecolabstract}
  \begin{center}
    \small
    \textbf{\keywordname}
    \\\medskip
    \mykeywords
  \end{center}
  \bigskip
  \onehalfspacing
\fi
\makeatother
}

\author{%
G. M. Cordeiro%
\thanks{%
G. M. Cordeiro
and
A. D. C. Nascimento
are with the
Departamento de Estat\'{\i}stica,
Universidade Federal de Pernambuco (UFPE), Brazil.
E-mail: \url{gauss,abraao@de.ufpe.br}}
\and
R.~J.~Cintra%
\thanks{%
R. J. Cintra is with the
Signal Processing Group,
CCEN,
Universidade Federal de Pernambuco (UFPE), Brazil.
E-mail: \url{rjdsc@de.ufpe.br}}
\and
L. C. R\^ego%
\thanks{%
L. C. R\^ego
is with the  Departamento de Estat\'{\i}stica e Matem\'atica Aplicada,
Universidade Federal do Cear\'a, Brazil.
E-mail: \url{leandro@dema.ufc.br}}
\and
A. D. C. Nascimento${}^\ast$%
}

\title{%
The Gamma Generalized Normal Distribution: A Descriptor of SAR Imagery}

\newcommand{\myabstract}{%
We propose a new four-parameter distribution for modeling synthetic aperture
radar (SAR) imagery named the gamma generalized normal (GGN) by combining the gamma and
ge\-ne\-ralized normal distributions. A mathematical characterization of the new distribution
is provided by identifying the limit behavior and by calculating the density and moment expansions.
The GGN model performance is evaluated on both synthetic and actual data and,
for that, maximum likelihood estimation and random number generation are discussed.
The proposed distribution is compared with the beta generalized normal distribution (BGN),
which has already shown to appropriately represent SAR imagery.
The performance of these two distributions are measured by means of
statistics which provide evidence that the GGN can outperform the BGN distribution in some contexts.
}

\newcommand{\mykeywords}{%
Gamma generalized distribution,
Generalized normal,
Maximum likelihood,
Moment,
SAR images.
}

\date{}

\begin{document}

\printtitle

\section{Introduction}
\label{GGN:1}

The statistics literature is filled with hundreds of continuous univariate distributions
and se\-veral recent developments focus on new techniques for building meaningful
distributions. Nume\-rous classical distributions have been extensively used over the last
decades for modeling data in se\-ve\-ral areas such as biological studies,
environmental sciences,
engineering, economics, actuarial sciences, finance,
de\-mo\-gra\-phy, in\-su\-ran\-ce and me\-di\-cal sciences.
However, in many applied areas
such as image processing, finance and insurance data, there is a clear need for
extended forms of these distributions.
Two core reasons are:
\begin{itemize}
\item[a)]
The proposal of extended models often
flexibilizes
the adjustment of
a model (say \emph{baseline})
that has a connection more direct with the phenomenon origin,
but it may not work well in practice.
Further,
the extensions
include
the baseline model
as particular case,
easily identified in the extended distribution parametric space.
This last fact
delivers
to the future users of the proposal
a simple rule
(based on asymptotic results from maximum likelihood estimators)
to choose between
the baseline or extended models
before actual scenarios.
\item[b)]
Theses extensions
also may stem from a phenomenological justification
and
introduce baselines
in a physical, industrial, and biological contexts:

\begin{itemize}

\item[(i)]
Time to relapse of cancer under the first-activation scheme~\cite{Vicente2013}.

\item[(ii)]
The failure of a device which occurs due to the presence of an unknown number of factors~\cite{ADAMIDIS199835,llai2013}.

\item[(iii)]
Time of resistance to a disease manifestation due to
the $i$th latent factor has the baseline distribution~\cite{SILVA2016119}.
\end{itemize}
\end{itemize}

The generalized normal (``\mbox{GN}'' for short) distribution with location parameter $\mu$, dispersion parameter
$\sigma$,
and shape parameter $s$ has probability density function (\mbox{pdf}) given by Nadarajah~\cite{nadarajah2005generalized}
\[
g(x)=
\frac{s}{2\sigma\,\Gamma(1/s)}
\exp\left\{-\left|\frac{x-\mu}{\sigma}\right|^s\right\},
\quad x\in \mathbb{R},
\]
where $\Gamma(a) = \int_{0}^{\infty}
t^{a-1}\,\rm{e}^{-t}dt$ is the gamma function,
$\mu$ is a real number, and $\sigma$ and $s$ are positive real numbers.
For the special case $s=1$, the above density function reduces to the Laplace
distribution with location parameter $\mu$ and scale parameter $\sigma$.
Similarly,
for $s=2$, the normal distribution is obtained with mean
$\mu$ and variance $\sigma^2/2$.
The main feature of the GN model is that
the new parameter $s$
can add
some skewness and kurtosis.

The cumulative distribution function (\mbox{cdf}) of the \mbox{GN} distribution can be expressed
as~\cite[Eq.~5-6]{nadarajah2005generalized}
\begin{equation}\label{eq.F(x)}
G(x)=
\begin{cases}
\frac{\Gamma\left(1/s, (\frac{\mu-x}{\sigma})^s\right)}{2\Gamma(1/s)},
& x \leq \mu,\\
&\\
1-\frac{\Gamma(1/s, \left(\frac{x-\mu}{\sigma}\right)^s)}{2\Gamma(1/s)},
& x > \mu,
\end{cases}
\end{equation}
where $\Gamma(s,x) = \int_x^\infty t^{s-1} \rm{e}^{-t}\, \mathrm{d}t$
is the upper
incomplete gamma function.

The density function of the standardized random variable
$Z=(X-\mu)/\sigma$
is
$$
\phi_s(z)=\frac{s}{2 \Gamma(1/s)}\exp(-|z|^s),\quad z\in \mathbb{R}.
$$
Thus, $g(x)=\sigma^{-1}\,\phi_s(\frac{x-\mu}{\sigma})$.
From Equation~\eqref{eq.F(x)}, the \mbox{cdf} of the standardized GN
distribution reduces to
\begin{align*}
\Phi_s(z)=
\begin{cases}
\frac{\Gamma(1/s,(-z)^s)}{2\Gamma(1/s)},& z \leq 0,\\
&\\
1-\frac{\Gamma(1/s, z^s)}{2\Gamma(1/s)},& z > 0.
\end{cases}
\end{align*}

A family of univariate distributions generated by gamma random variables
was pioneered by
Zografos and Balakrishnan~\cite{Zografos2009344}
and
Risti\'c and N. Balakrishnan~\cite{RistBalakrishnan}.
They
defined
the {\it gamma generalized-G} (``\mbox{GG-G}'' for short) distribution
from any baseline \mbox{cdf} $G(x)$, $x \in\mathbb{R}$,
using an additional shape parameter $a>0$.
The associated \mbox{pdf} and \mbox{cdf} are given by
\begin{eqnarray}\label{pdf:ZB}
f(x)=\frac{g(x)}{\Gamma(a)}\,\left\{-\log\left[1-G(x)\right]\right\}^{a-1}
\end{eqnarray}
and
\begin{align}\label{cdf:ZB}
F(x)=\frac{1}{\Gamma(a)}\,\int_{0}^{-\log\left[1-G(x)\right]}\,t^{a-1}\,\rm{e}^{-t}\,\mathrm{d}t
=\gamma_1(a,-\log\left[1-G(x)\right]),
\end{align}
respectively, where $g(x) = \mathrm{d}G(x)/\mathrm{d}x$,
$\gamma(a,z) = \int_{0}^{z} t^{a-1}\,\rm{e}^{-t}dt$
is
the lower
incomplete gamma function
and $\gamma_1(a,z)=\gamma(a,z)/\Gamma(a)$
is
the lower incomplete gamma function ratio.

Each new \mbox{GG-G} distribution can be obtained from a specified G distribution.
For $a = 1$,
the \mbox{G} distribution is a basic exemplar with a continuous crossover towards cases with different shapes (for example,
a particular combination of skewness and kurtosis).
Zografos and Balakrishnan~\cite{Zografos2009344} motivated the GG distribution as follows.
Let $X_{(1)},\ldots, X_{(n)}$ be lower record values from a sequence of independent and identically distributed (i.i.d.) random variables from a population
with \mbox{pdf} $g(x)$.
Then, the \mbox{pdf} of the $n$th lower record value is given by~\eqref{pdf:ZB} with $a=n$.
A logarithmic transformation of the baseline distribution G transforms the random variable $X$ with
density function~\eqref{pdf:ZB} to a gamma distribution.
In other words, if $X$ has the density~\eqref{pdf:ZB},
then $Z=-\log[1-G(X)]\sim$Gamma$(a,1)$ has \mbox{pdf} expressed as
$$
\pi(z;a)=\Gamma(a)^{-1}\,z^{a-1}\,{\rm e}^{-z},
$$
where $z\in\mathbb{R}_+$.
The opposite is also true, if $Z\sim$Gamma$(a,1)$, then the random variable
$X=G^{-1}(\,1-\exp[-Z]\,)$ has the \mbox{GG-G} density function~\eqref{pdf:ZB}.
Nadarajah~\emph{et al.}~\cite{NadarajahCordeiroOrtega}
derived some  mathematical properties of \eqref{pdf:ZB} in the most simple,
explicit, and general forms for any G distribution.

The first goal of this paper is to develop an extension of the generalized normal (GN) distribution defined from~\eqref{pdf:ZB}:
the \textit{gamma generalized normal} (``GGN'' for short) distribution. It may be mentioned that although several skewed distributions
exist on the positive real axis, not many skewed distributions
are available on the whole real line, which are easy to use for data analysis purposes.

The main
role of the extra parameter $a>0$ is that the GGN distribution can be used to model skewed real data,
a feature very common in practice.
This distribution with four parameters can control location,
dispersion,
and skewness with great flexibility.

As a second goal, we
advance the GGN distribution as a model for
the image processing of SAR imagery~\cite{ElZaartZiou2007}.
We
provide
a data analysis using simulated and actual SAR imagery.
Also,
we compare the GGN model with
the beta generalized normal (BGN) distribution introduced by Cintra \emph{et al.}~\cite{cintraetal2013}.
The BGN law has outperformed several existing models used for fitting SAR images,
such as the gamma distribution~\cite{Delignonetal2002},
the $\mathcal{K}$ distribution~\cite{Blacknell1994}, and
the $\mathcal{G}^0$ distribution~\cite{freryetal1997a}.
The gamma distribution is regarded as standard model for the herein considered types of SAR data~\cite{Delignonetal2002}.
For such, actual data is analyzed and fitted according to GGN and BGN models above by means of eight figures of merit~\cite{Gao2010}.
The model we propose
outperforms
the BGN distribution.
In particular,
the BGN
has failed to describe
regions having more than one texture for all considered channels,
the GBN works well.

The rest of the paper
is organized
as follows.
In Section~\ref{GGN:2},
we define the GGN distribution,
derive its density,
discuss special cases
and provide some of its characteristics.
A useful expansion for its density function
is determined in Section~\ref{GGN:3}.
Explicit expressions for the moments
are derived
in Section~\ref{GGN:4}.
Maximum likelihood estimation of the model parameters
is investigated
in Section~\ref{GGN:6}.
In Section~\ref{GGN:7},
we
propose
a random number generator for the new distribution.
Section~\ref{GGN:8}
details
the SAR image analysis based on real data.
Section~\ref{GGN:9}
provides
concluding remarks.

\section{The GGN Distribution}
\label{GGN:2}

Based on Equations~\eqref{pdf:ZB} and~\eqref{cdf:ZB},
we propose a generalization
of the \mbox{GN} distribution and provide a comprehensive treatment of its
mathematical properties.
The \mbox{pdf} and \mbox{cdf} of the \mbox{GGN} distribution are given by
\begin{equation}
\label{ggndensity}
f(x)=\frac{\phi_s\left(\frac{x-\mu}{\sigma}\right)}{\sigma\,\Gamma(a)}\,\left\{-\log\left[1-\Phi_s\left(\frac{x-\mu}{\sigma}\right)\right]\right\}^{a-1}
\end{equation}
and
\begin{align*}%
F(x)=\frac{1}{\Gamma(a)}\,\int_{0}^{-\log\left[1-\Phi_s\left(\frac{x-\mu}{\sigma}\right)\right]}\,t^{a-1}\,\rm{e}^{-t}\,\mathrm{d}t%
=\gamma_{1}\left(a,-\log\left[1-\Phi_s\left(\frac{x-\mu}{\sigma}\right)\right]\right).
\end{align*}
Here, the parameter~$a$ affects the skewness through the
relative tail weights. It provides greater flexibility in the form of the
distribution and consequently in modeling observed real data.
Hereafter, a random variable $X$ with density function~\eqref{ggndensity} is denoted by
$X\sim \text{GGN}(\mu,\sigma,s,a)$.
Notice that the gamma normal and gamma Laplace distributions are special cases
of (\ref{ggndensity}) for $s=1$ and $s=2$, respectively.
The baseline GN distribution is also a sub-model
for $a=1$.
The normal distribution
arises for $a=1$ and $s=2$, whereas the Laplace distribution corresponds
to $a=s=1$.
A noteworthy property of~\eqref{ggndensity}
is its ability for fitting skewed real data that cannot be
properly fitted by existing distributions with support on the real line. Location and scale
para\-me\-ters
can be set to zero and one, respectively,
in \eqref{ggndensity}.
This is because
if $W\sim\operatorname{GGN}(0,1,s,a)$
then $X=\sigma\,W+\mu \sim\operatorname{GGN}(\mu,\sigma,s,a)$.

The GGN distribution has two important interpretations.
First, given a sequence of independent and identically GN random variables with cdf \eqref{eq.F(x)},
the GGN density function coincides with the density function of the $a$th lower record value if $a$ is a positive integer.
As a consequence,
\eqref{ggndensity} can be considered as a generalization of the density function of GN lower records.
Record va\-lues arise naturally in many real life problems relating to engineering, economics and medicine.
Secondly, if $Z$ has a gamma density with scale parameter one and shape parameter $a$, the GN quantile function
evaluated at the random variable $1-\mathrm{e}^{-z}$ has pdf given by
\eqref{ggndensity}.

\begin{figure}
\centering

\subfigure[$a=1/2$]{\includegraphics[width=.48\linewidth]{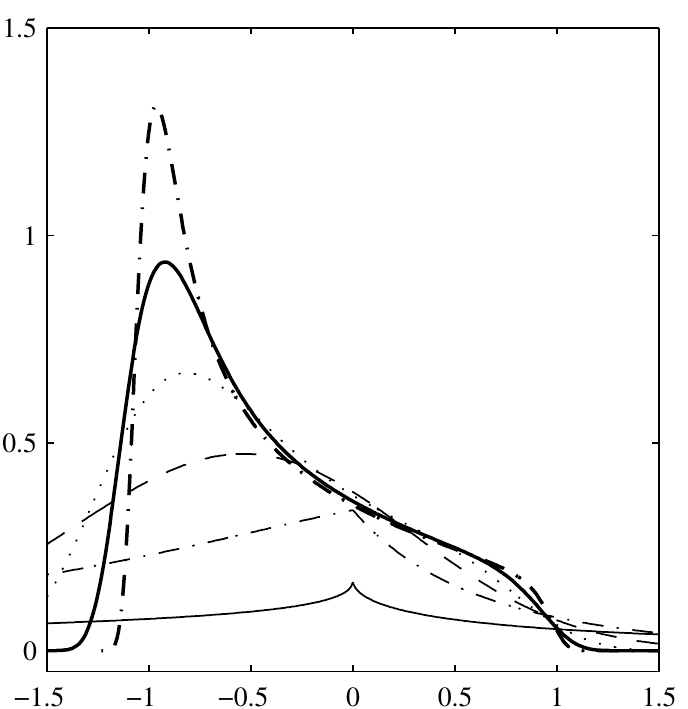}}
\subfigure[$a=3/4$]{\includegraphics[width=.48\linewidth]{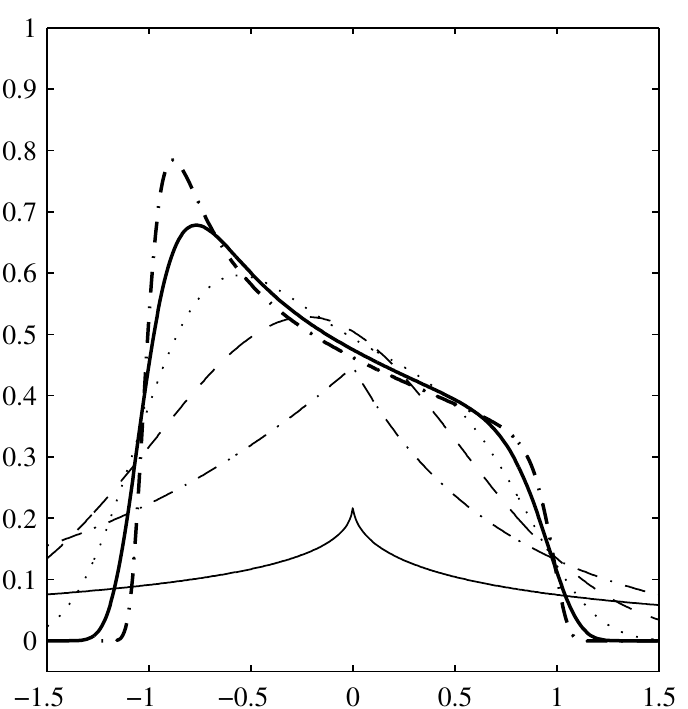}}
\\
\subfigure[$a=1$]{\includegraphics[width=.48\linewidth]{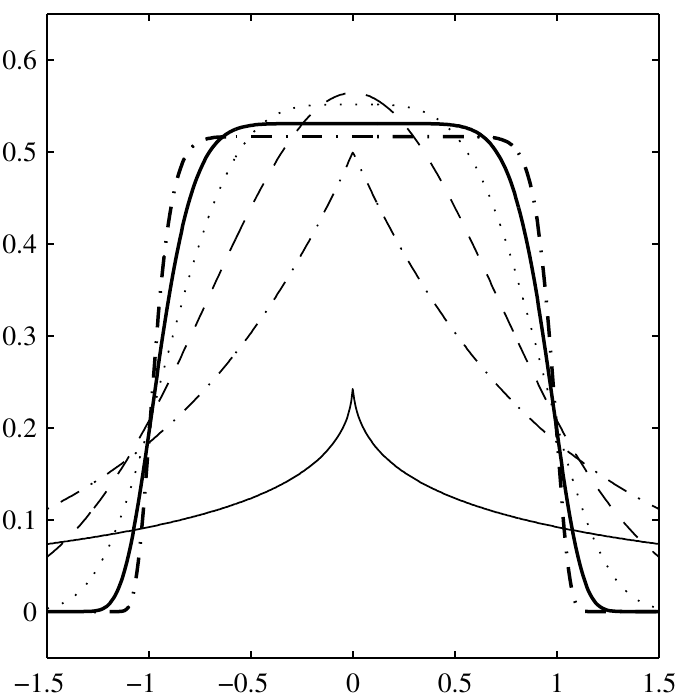}}
\subfigure[$a=2$]{\includegraphics[width=.48\linewidth]{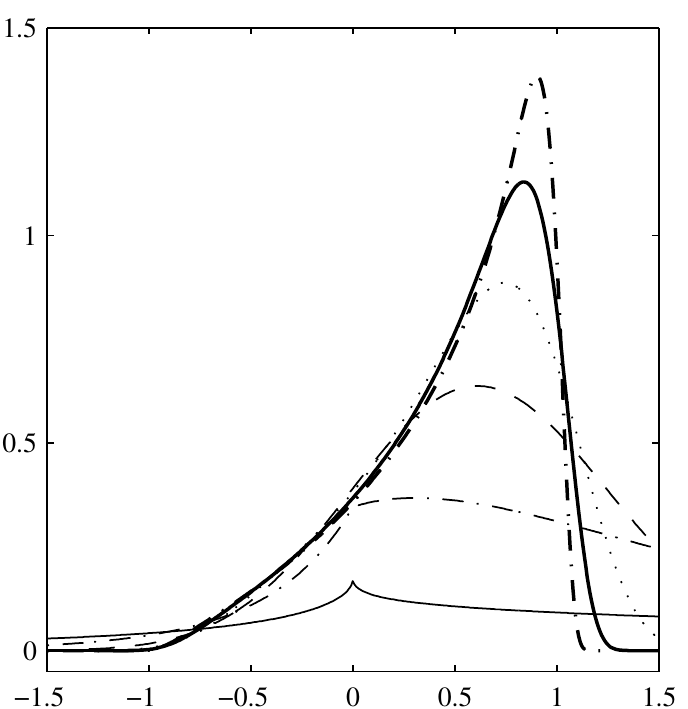}}

\caption{%
Plots of the GGN density function for some parameter values:
$\mu=0$, $\sigma=1$,
and
{%
$s\in\{16, 8 , 4 , 2, 1, 1/2 \}$
(dash-dotted, solid, dotted, dashed, dash-dotted, solid).
}
}
\label{figure-plots-pdf}
\end{figure}

Note that the limiting behavior of the GN density with respect to~$s$
is given by
\begin{align*}
\lim_{s\to\infty}
\phi_s(z)=
\lim_{s\to\infty}
\frac{s}{2 \Gamma(1/s)}
\exp
(-|z|^s)=
\begin{cases}
\frac{1}{2}, & |z|<1, \\
0, & \text{otherwise.}
\end{cases}
\end{align*}
Consequently,
by invoking the dominate convergence theorem,
we obtain:
\begin{align*}
\lim_{s\to\infty}
\Phi_s(z)
=
\begin{cases}
0, & z<-1, \\
\frac{1}{2}(z+1), & |z|<1,\\
1, &z>1.
\end{cases}
\end{align*}
Making appropriate substitutions,
the limiting distribution $\lim_{s\to\infty}f(x)$
is given by
\begin{align*}
\label{limit}
\lim_{s\to\infty}
f(x)
&=
\frac{1}{2\,\sigma\,\Gamma(a)}
\left\{
-\log
\left[
1 -
\left(
\frac{1}{2} + \frac{x-\mu}{2\sigma}
\right)
\right]
\right\}^{a-1}
\\
&=
f_\text{GU}
\left(
\frac{x-\mu}{\sigma}
\right)
,
\quad
x\in[\mu-\sigma,\mu+\sigma],
\end{align*}
where
$f_\text{GU}(x)$
is
the pdf of the gamma uniform random variable
associated to the uniform distribution
over the interval $[-1,1]$.
Figure~\ref{figure-plots-pdf} illustrates this behavior.

\section{Density expansion}
\label{GGN:3}

Expansions for~\eqref{pdf:ZB} and \eqref{cdf:ZB} can be derived using the concept of exponentiated distributions.
Consider the \textit{exponentiated generalized normal} (``EGN'') distribution with
power parameter $c>0$ defined by $Y_c \sim \mbox{EGN}(\mu,\sigma,s,c)$ having cdf and pdf given by
\begin{equation}
\label{cdfEGN}
H_c(x)=
\left\{
\begin{array}{l}
\left[\frac{\Gamma\left(1/s, (\frac{\mu-x}{\sigma})^s\right)}{2\Gamma(1/s)}\right]^c,\,x \leq \mu,\\%\nonumber\\
\left[1-\frac{\Gamma(1/s, \left(\frac{x-\mu}{\sigma}\right)^s)}{2\Gamma(1/s)}\right]^c,\,x > \mu,
\end{array}
\right.
\end{equation}
and
\begin{equation}\label{pdfEGN}
h_c(x)=%
\left\{
\begin{array}{l}
\frac{s}{2\sigma\,\Gamma(1/s)}\,\exp\left\{-\left|\frac{x-\mu}{\sigma}\right|^s\right\}\,\left[\frac{\Gamma\left(1/s, (\frac{\mu-x}{\sigma})^s\right)}{2\Gamma(1/s)}\right]^{c-1},\,%
\quad x \leq \mu,\\%\nonumber\\
\frac{s}{2\sigma\,\Gamma(1/s)}
\exp\left\{-\left|\frac{x-\mu}{\sigma}\right|^s\right\}\,\left[1-\frac{\Gamma(1/s, \left(\frac{x-\mu}{\sigma}\right)^s)}{2\Gamma(1/s)}\right]^{c-1},%
\quad x > \mu,
\end{array}
\right.
\end{equation}
respectively.
The properties of some exponentiated distributions
have been stu\-died by several authors,
see
Mudholkar and Srivastava~\cite{MudholkarSrivastava1993},
Mudholkar \emph{et al.}~\cite{Mudholkaretal1995} for the exponentiated Weibull (EW),
Gupta \emph{et al.}~\cite{Gupta1998} for the exponentiated Pareto,
Gupta and Kundu~\cite{GuptaandKundu} for the exponentiated exponential (EE),
Nadarajah and Gupta~\cite{NadarajahandGupta} for exponentiated gamma (EG) distributions,
and
Cordeiro \emph{et al.}~\cite{Cordeiro2011kk}
for
the exponentiated generalized gamma (EGG) distribution.

Based on an expansion due to
Nadarajah \emph{et al.}~\cite{NadarajahCordeiroOrtega},
we can write
\begin{align*}
\Bigg\{-\log\left[1-\Phi_s\left(\frac{y-\mu}{\sigma}\right)\right]\Bigg\}^{a-1}%
\!\!\!\!\!=&(a-1)\sum_{k=0}^{\infty}{k+1-a \choose k}\,
\sum_{j=0}^{k}\frac{(-1)^{j+k}{k \choose j}\,p_{j,k}}{(a-1-j)}\nonumber\\
&\times\,\Phi_s\left(\frac{y-\mu}{\sigma}\right)^{a+k-1},
\end{align*}
where $a > 0$ is any real parameter and the constants $p_{j,k}$ can be calculated recursively by
\begin{eqnarray*}%
p_{j,k} = k^{-1} \sum_{m=1}^{k}\,\frac{(-1)^{m}\,[m(j+1)-k]}{(m+1)}\,p_{j,k-m},
\end{eqnarray*}
for $k = 1, 2, \ldots$ and $p_{j,0} = 1$. Let
\begin{eqnarray*}
b_{k}= \frac{{k+1-a \choose k}}{(a+k)\Gamma(a-1)}\,\sum_{j=0}^{k}\frac{(-1)^{j+k}\,{k \choose j}\,\,p_{j,k}}{(a-1-j)}.
\end{eqnarray*}
Then, \eqref{ggndensity} can be expressed as
\begin{eqnarray}\label{pdf_ggn_exp}
f(x)=\sum_{k=0}^{\infty}\,b_{k}\,h_{a+k}(x),
\end{eqnarray}
where $h_{a+k}(x)$ denotes the density function
of the
random variable
$Y_{a+k}\sim\mbox{EGN}(a + k,\mu,\sigma,s)$ given by \eqref{pdfEGN}.
The cdf corresponding to (\ref{pdf_ggn_exp}) becomes
\begin{eqnarray}\label{cdf_ggn_exp}
F(x)=\sum_{k=0}^{\infty}\,b_{k}\,H_{a+k}(x),
\end{eqnarray}
where $H_{a+k}(x)$ denotes the cumulative function of $Y_{a+k}$ given by (\ref{cdfEGN}).

\section{Moments}
\label{GGN:4}

Let $X$ be a random variable having the GGN$(\mu,\sigma,s,a)$
distribution.
The $n$th moment of $X$ becomes
\begin{align}\label{mom1}
\mathrm{E}(X^n)=&\sum_{k=0}^{\infty}\,\frac{(a+k)\,b_{k}}{\sigma} %
\,\int_{-\infty}^\infty x^n\,
\left[\Phi_s\left(\frac{x-\mu}{\sigma}\right)\right]^{a+k-1}
\phi_s \left(\frac{x-\mu}{\sigma}\right) \mathrm{d}x.
\end{align}
Defining the auxiliary function
$h_k^{(s)}(x)=
\Phi_s\left(
\frac{x-\mu}{\sigma}
\right)^{a+k-1}
\phi_s\left(\frac{x-\mu}{\sigma}\right)$ and setting $z =\frac{x-\mu}{\sigma}$, we can write
\begin{align}
\nonumber
\int_{-\infty}^\infty x^n\,h_k^{(s)}(x)
\mathrm{d}x&=
\sigma\,\int_{-\infty}^\infty
(\sigma z+\mu)^n\,\Phi_s(z)^{a+k-1}\,\phi_s(z)
\mathrm{d}z\nonumber \\
&=
\sigma
\mu^n\,\sum_{i=0}^n\,\binom{n}{i}\,
\left(\frac{\sigma}{\mu}
\right)^i\,\int_{-\infty}^\infty z^i\,
\Phi_s(z)^{a+k-1}\,\phi_s(z)\mathrm{d}z.
\label{eq.section5}
\end{align}

Splitting the integration range in~\eqref{eq.section5} in two parts
and considering that $\Phi_s(-z)= 1-\Phi_s(z)$ and
$\phi_s(z) =\phi_s(-z)$, we have
\begin{align*}
\int_{-\infty}^\infty z^i\,\phi_s(z)\,\Phi_s(z)^{a+k-1}\mathrm{d}z
=&\int_0^\infty z^i\,\phi_s(z)\,\Phi_s(z)^{a+k-1}\mathrm{d}z
\\
&+ \int_0^\infty (-1)^i\,z^i\,\phi_s(z)
[1-\Phi_s(z)]^{a+k-1}\mathrm{d}z.
\end{align*}

We can express $\Phi_s(z)^\alpha$ for any real $\alpha>0$ in terms of a power series of $\Phi_s(z)$,
namely $\Phi_s(z)^\alpha=\sum_{j=0}^\infty v_j(\alpha)\,\Phi_s(z)^j$,
where
$v_j(\alpha)=\sum_{m=j}^\infty (-1)^{j+m}\,\binom{\alpha}{m}\,\binom{m}{j}$.
Thus, based on this expansion, we obtain
\begin{align*}
&\int_{-\infty}^\infty z^i\,\phi_s(z)\,\Phi_s(z)^{a+k-1}
\mathrm{d}z=\\
&\sum_{j=0}^\infty v_j(a+k-1)\,\int_0^\infty z^i\,\phi_s(z)\,\Phi_s(z)^j
\mathrm{d}z\\
&+\sum_{j=0}^\infty (-1)^{i+j}\,\binom{a+k-1}{j}\,\int_0^\infty
z^i\,\phi_s(z)\,\Phi_s(z)^j\mathrm{d}z.
\end{align*}

We define the $(i,j)$th probability weighted moment of $Z$ by
$J_{i,j}^{(s)}=\int_0^\infty z^i\,\phi_s(z)\,\Phi_s(z)^j \mathrm{d}z
$,
where $i,j$ are positive integers.
Cintra \emph{et al.}~\cite{cintraetal2013} de\-mons\-tra\-ted that
\begin{align}
\label{J-form-1}
J_{i,j}^{(s)}
=&
\frac{1}{[2\Gamma(1/s)]^{j+1}}
\sum_{r=0}^j\,\binom{j}{r}\,\Gamma(s^{-1})^{j-r}\nonumber \\
&\times \sum_{m=0}^\infty c_{m,r}\,
\Gamma\left(m+\frac{i+r+1}{s}\right),
\end{align}
where $c_{0,r}=s^r$ and for all $m \geq 1$
$$c_{m,r}=(m\,s)^{-1}\sum_{\ell=1}^{m}\frac{(-1)^{\ell}\,[(r+1)\ell-m]}{(s^{-1}+\ell)\ell!}\,c_{m-\ell,r}.$$
Hence,
we obtain:
\begin{align}\label{mom2}
&\int_{-\infty}^\infty
z^i\,\phi_s(z)\,\Phi_s(z)^{a+k-1}\,\mathrm{d}z
=\nonumber \\
&\sum_{j=0}^\infty
\left[v_j(a+k-1)+(-1)^{i+j}\,\binom{a+k-1}{j}\right]\,J_{i,j}^{(s)}.
\end{align}
Finally,
based on \eqref{mom1}, \eqref{eq.section5} and \eqref{mom2},
we have the following expression:
\begin{align}\label{mom3}
&\mathrm{E}(X^n)=\sum_{k,j=0}^{\infty}\,(a+k)\,b_{k}\,\sum_{i=0}^n\,\binom{n}{i}\,
\frac{\sigma}{\mu^{i-n}}\,\nonumber \\
&\quad\times \left[v_j(a+k-1)+(-1)^{i+j}\,\binom{a+k-1}{j}\right]\,J_{i,j}^{(s)}.
\end{align}
The above expression is the main result of this section.

\section{Maximum Likelihood Estimation}
\label{GGN:6}

Consider
a random variable $X$ having the GGN distribution
and
let
$\bm{\theta}=(\,\mu,\,\sigma,\,s,\,a\,)^\top$
be the model parameters.
Thus,
the associated log-likelihood function
for one observation $x$ is
\begin{align}
\ell(\bm{\theta})\,=&\,\ell(\bm{\theta};x)\,=\,\log\,\Big[\phi_s\Big(\frac{x-\mu}{\sigma}\Big)\Big]\,-\,\log\,(\sigma)\,-\,\log[\Gamma(a)]\nonumber \\
\,&+\,(a-1)\,\log\Big\{\,-\,\log\,\Big[\,1\,-\,\Phi_s\Big(\,\frac{x-\mu}{\sigma}\,\Big)\,\Big]\,\Big\}.
\label{Sugg:Lean}
\end{align}
The maximum likelihood estimate (MLE) of $\bm{\theta}$
is determined by
maximizing
$
\ell_n(\bm{\theta})\,=\,\sum_{i=1}^n\,\ell(\bm{\theta};x_i)
$
from a dataset $x_1,x_2,\ldots,x_n$.

Based on Equation~\eqref{Sugg:Lean},
the score function can be derived as follows
\begin{align*}
\bm{U}_\theta\,&=\,(\,U_\mu,\,U_\sigma,\,U_s,\,U_a\,)^\top\\
\,&=\,
\Big(\,\frac{\mathrm{d}\,\ell(\bm{\theta})}{\mathrm{d}\mu},
\,\,\,\frac{\mathrm{d}\,\ell(\bm{\theta})}{\mathrm{d}\sigma},
\,\,\,\frac{\mathrm{d}\,\ell(\bm{\theta})}{\mathrm{d}s},
\,\,\,\frac{\mathrm{d}\,\ell(\bm{\theta})}{\mathrm{d}a}\Big)^\top.
\end{align*}

To that end,
we
consider initially
the discussion of some auxiliary results.
After some algebra,
two identities hold:
\begin{equation}
\frac{\mathrm{d}\,\Phi_s\,\Big(\,\frac{x-\mu}\sigma\,\Big)}{\mathrm{d}\,\delta}\,=\,C_\delta\,\phi_s\,\Big(\,\frac{x-\mu}\sigma\,\Big)
\label{GGN:6:eq:1}
\end{equation}
and
\begin{equation}
\frac{\mathrm{d}\,\log\,\phi_s\,\Big(\,\frac{x-\mu}\sigma\,\Big)}{\mathrm{d}\,\delta}\,=\,-\,s\,C_\delta\,\operatorname{sign}(\,x\,-\,\mu\,)\,\Big|\,\frac{x-\mu}\sigma\,\Big|^{s-1},
\label{GGN:6:eq:2}
\end{equation}
where $\delta\in\{\mu,\sigma\}$,
$C_\mu=-\sigma^{-1}$
and
$C_\sigma=\mu/\sigma^2$.
Another result required for the subsequent derivations is
$\frac{\mathrm{d}}{\mathrm{d}s}\Phi_s\left(\frac{x-\mu}{\sigma}\right),$
which
depends
on the derivatives of the gamma and the incomplete gamma functions.
Cintra \emph{et al.}~\cite{cintraetal2013} demonstrated
that identities~\eqref{GGN:6:eq:1} and~\eqref{GGN:6:eq:2} for $\delta=s$ are given by
\begin{align*}
&\frac{\mathrm{d}\,\Phi_s\left(\frac{x-\mu}{\sigma}\right)}{\mathrm{d}s}
=
\frac{1}{2\Gamma(1/s)}\\
&\times
\begin{cases}
s^2\tilde\psi\left( s, (\frac{\mu-x}{\sigma})^s\right)+ \psi(1/s)\Gamma(1/s, (\frac{\mu-x}{\sigma})^s),
&
x\leq\mu,\\
&\\
-[s^2\tilde\psi\left( s, (\frac{x-\mu}{\sigma})^s\right)+ \psi(1/s)\Gamma(1/s, (\frac{x-\mu}{\sigma})^s)],
&
x>\mu,
\end{cases}
\end{align*}
where
$
\psi(x)
$
is the digamma function,
and
$
\tilde\psi(s,x)
\triangleq
\frac{\mathrm{d}\,\Gamma(1/s,x^s)}{\mathrm{d}s}
$,
for $x>0$~\cite{cintraetal2013}.
Additionally, the following result
was also presented
by Cintra \textit{et al.}:
\begin{align*}
\frac{\mathrm{d}\,\log\phi_s\left(\frac{x-\mu}{\sigma}\right)}{\mathrm{d}s}
&=
\frac{1}{s}
+
\frac{\psi(1/s)}{s^2}
-
\left|\frac{x-\mu}{\sigma} \right|^s
\log
\left(
\left|\frac{x-\mu}{\sigma} \right|
\right).
\end{align*}
The elements of the GGN score function are given in Fig.~\ref{fig:scores}.

\begin{figure*}[hbt]
\hrulefill
\begin{gather}
U_\mu\,=\,\frac s{\sigma}\,\operatorname{sign}(\,x\,-\,\mu\,)\,\Big|\,\frac{x-\mu}\sigma\,\Big|^{s-1}\,-\,\Big(\frac{a-1}{\sigma}
\Big)\,\frac{\phi_s\,\Big(\,\frac{x-\mu}\sigma\,\Big)}{[1-\Phi_s\,\Big(\,\frac{x-\mu}\sigma\,\Big)]\,\log[1-\Phi_s\,\Big(\,\frac{x-\mu}\sigma\,\Big)]},
\\
U_\sigma\,=\,-\frac {s\mu}{\sigma^2}\,\operatorname{sign}(\,x\,-\,\mu\,)\,\Big|\,\frac{x-\mu}\sigma\,\Big|^{s-1}\,-\,\frac 1\sigma\,+\,\Big[\frac{(a-1)\mu}{\sigma^2}
\Big]\,\frac{\phi_s\,\Big(\,\frac{x-\mu}\sigma\,\Big)}{[1-\Phi_s\,\Big(\,\frac{x-\mu}\sigma\,\Big)]\,\log[1-\Phi_s\,\Big(\,\frac{x-\mu}\sigma\,\Big)]},
\\
U_s\,=\,\frac {\mathrm{d}\,\log\,\phi_s\,\Big(\,\frac{x-\mu}\sigma\,\Big)}{\mathrm{d} s}\,+\,(a-1)\,\frac{
\frac {\mathrm{d}\,\Phi_s\,\Big(\,\frac{x-\mu}\sigma\,\Big)}{\mathrm{d} s}
}{[1-\Phi_s\,\Big(\,\frac{x-\mu}\sigma\,\Big)]\,\log[1-\Phi_s\,\Big(\,\frac{x-\mu}\sigma\,\Big)]}
\\
U_a\,=\,\,\log\Big\{\,-\,\log\,\Big[\,1\,-\,\Phi_s\Big(\,\frac{x-\mu}{\sigma}\,\Big)\,\Big]\,\Big\}\,-\,\psi(\,a\,).
\end{gather}
\hrulefill
\caption{The GGN score function.}\label{fig:scores}
\end{figure*}

\section{GGN Random Number Generator}\label{GGN:7}

Here, we give a random number generator (RNG) for the proposed distribution.
This RNG
provides
a Monte Carlo study
to determine
the influence of the GGN shape parameters $s$ and~$a$.

Among the algorithms
for generating continuous random variables,
we
use
the inverse transform due to its
analytical tractability~{\cite[p.~67]{ross2006simulation}}.
Let $X\sim \text{GGN}(\mu,\sigma,s,a)$.
Then,
as discussed in Section~\ref{GGN:1},
$X$ can be associated with $Z\sim\operatorname{Gamma}(a,1)$
according to the identity:
$$
X\,=\,G^{-1}(\,1-\mathrm{e}^{-Z}]\,).
$$
Notice that
\begin{align*}
G(x)
=
\begin{cases}
\frac{1}{2}
\left\{
1 - F_\Gamma\left[\left(-\frac{x-\mu}{\sigma}\right)^s\right]
\right\}
,
&x\leq\mu, \\
&\\
\frac{1}{2}
\left\{
1 + F_\Gamma\left[\left(\frac{x-\mu}{\sigma}\right)^s\right]
\right\}
,
&x>\mu,
\end{cases}
\end{align*}
where $F_\Gamma(\cdot)$ is the gamma cdf
with shape and
scale parameters given by $1/s$ and one,
respectively.

Let $x$ and $z$ be realizations of $X$ and $Z$, respectively.
For $x\leq\mu$,
we obtain the inverse transformation according to
the following equation
\begin{align*}
\frac{1}{2}
\left\{
1 - F_\Gamma\left[\left(-\frac{x-\mu}{\sigma}\right)^s\right]
\right\}
=
1\,-\,\mathrm{e}^{-z}.
\end{align*}
Therefore,
\begin{align*}
x
=
\mu
-
\sigma
\left[F^{-1}_\Gamma(\,2\,\mathrm{e}^{-z}\,-\,1\,)\right]^{1/s}
,
\end{align*}
where $F^{-1}_\Gamma(\cdot)$ is the quantile function of the
gamma distribution with parameters $1/s$ and one.
Since
$0\leq F_\Gamma\left[\left(-\frac{x-\mu}{\sigma}\right)^s\right]\leq1$,
we have $0 \leq z \leq \log 2$.

Analogously,
for $x>\mu$,
we obtain:
\begin{align*}
x
=
\mu
+
\sigma
\left[
F^{-1}_\Gamma(\,1\,-\,2\,\mathrm{e}^{-z}\,)
\right]^{1/s}
,
\quad
\log\, 2 < z <\infty
.
\end{align*}

Thus,
the generation of $X$ can follow the algorithm:
\begin{algorithmic}[1]
\State Generate $Z\sim \Gamma(a,1)$
\If{$Z\in[0,\log 2]$}
   \State $X=\mu-\sigma \left[F^{-1}_\Gamma(\,2\,\mathrm{e}^{-Z}-\,1\,)\right]^{1/s}$
\Else
   \State $X=\mu+\sigma \left[F^{-1}_\Gamma(\,1\,-\,2\,\mathrm{e}^{-Z}\,)\right]^{1/s}$
\EndIf
\State \textbf{return} $X$.
\end{algorithmic}

\vskip+2ex

\begin{figure}
\centering
\subfigure[GGN($0,1,s,5$)\label{powerr1}]{\includegraphics[width=4.9cm,height=7cm]{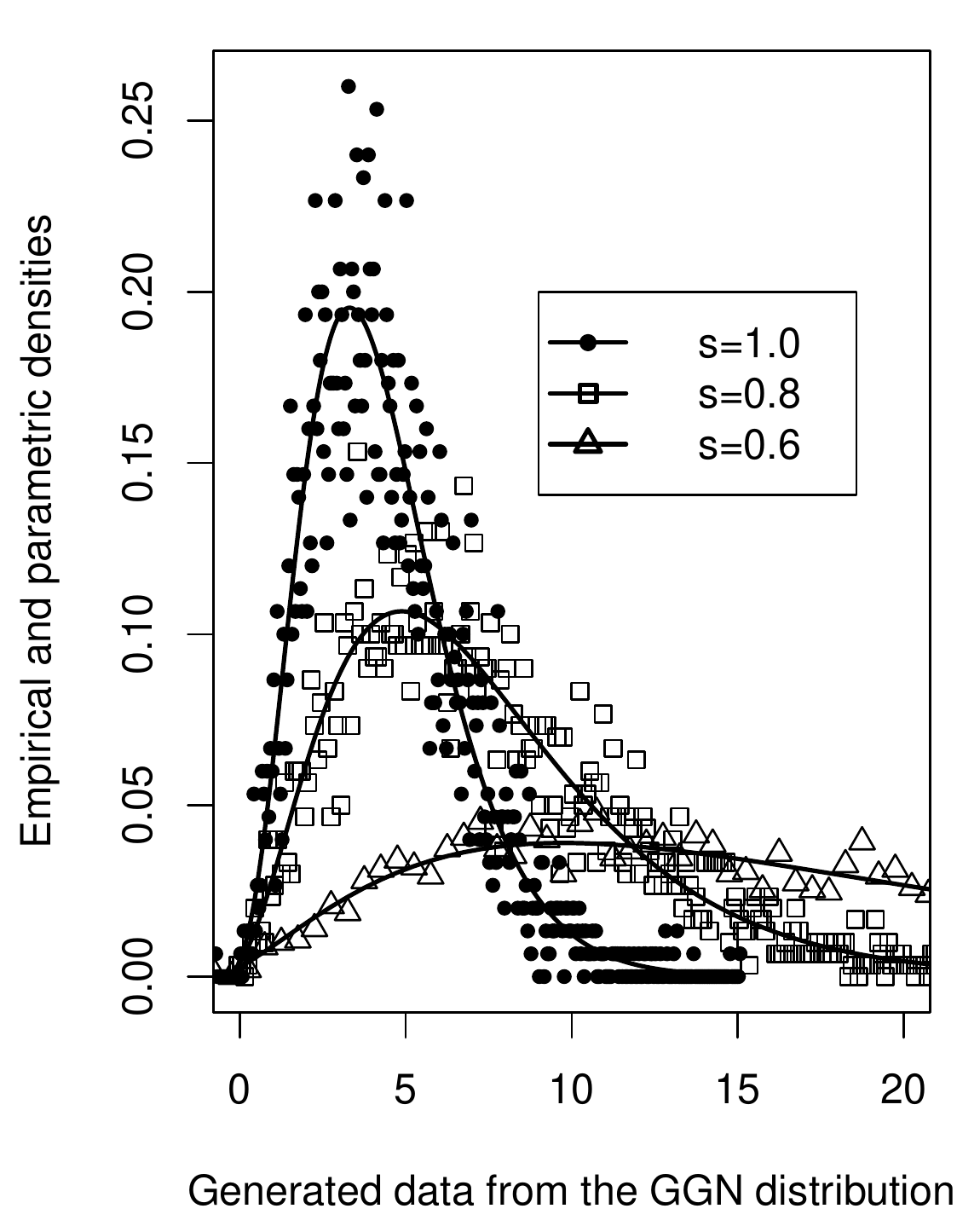}}
\subfigure[GGN($0,1,0.5,a$)\label{powerr2}]{\includegraphics[width=4.9cm,height=7cm]{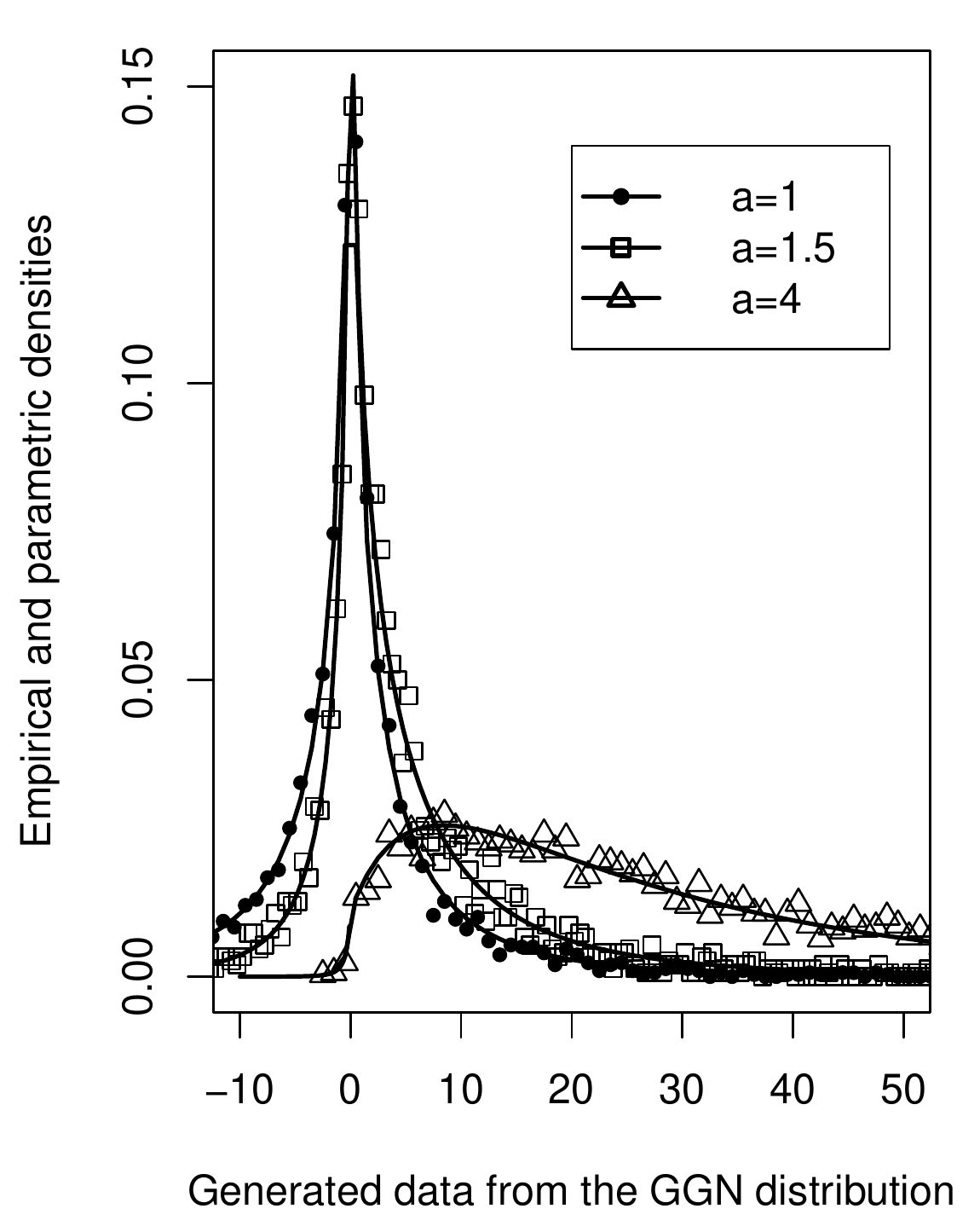}}
\caption{GGN random number generation.}
\label{passa0}
\end{figure}

Figure~\ref{passa0}
provides
six cases clustered on two graphics
where theoretical curves are compared with
randomly generated points.
The GGN law at
$\bm{\theta}=(0,\,1,\,s,\,5)^\top$
such that
$s\in\{0.6,\,0.8,\,1.0\}$
is considered in
Figure~\ref{powerr1}.
Data and density curves from
$X\sim \text{GGN}(\,0,\,1,\,0.5,\,a)$
such that $a\in\{1,\,1.5,\,4\}$
are displayed in
Figure~\ref{powerr2}.

\section{SAR Image Processing}\label{GGN:8}

In this section,
we assess
the GGN model performance on both
synthetic and actual data.
Using Monte Carlo experiments, we first
employ the proposed GGN RNG method to generate
synthetic data
in order to quantify
the influence of shape parameters $a$ and $s$
on the asymptotic properties of the MLEs.
Subsequently,
we
apply
the GGN model to actual SAR data,
which
require a specialized model
due to the presence of
the speckled noise in SAR imagery.

\subsection{Parameter study}

In order to quantify the influence of the shape parameters ($a$ and $s$)
on the MLEs from the proposed distribution and
on their corresponding mean square errors (MSEs), we
perform $M=5500$ Monte Carlo replications.
This simulation study follows the scheme:

\begin{enumerate}%
\item
Synthetic data
were generated
from the GGN distribution,
simulating ima\-ges of $5\times 5$, $7 \times 7$, and $11 \times 11$ pixels.
Therefore,
we
have
samples sizes of $N\in\{25,49,121\}$.

\item
By fixing the parameters $\mu=0$ and $\sigma=1$,
nine scenarios are considered
for
$a\in\{2,3\}$ and
$s\in\{1,2,3\}$.

\item
As assessing criteria,
the average of the MLEs and
their corresponding estimates for the MSEs
are computed.

\end{enumerate}

Table~\ref{table:resul:1}
presents
the numerical values for the selected criteria.
In general terms,
as expected,
the results
indicate
that
increasing the sample sizes
decreases the MSE
and
the estimates
become
closer to the true parameters in average.
In addition,
the estimates for $s$
assume
the worse values among them; i.e.,
the estimates of the MSE corresponding to $s$
converge
more slowly to zero.

\begin{table*}[htbp]

\centering

\footnotesize

\caption{Average of the estimated parameters and their respective estimates for the mean square errors}
\label{table:resul:1}

\begin{tabular}{c c c c c}

\toprule

$\quad N\quad$ &
\multicolumn{4}{c}{ \textit{Estimated parameters (MSE)}} \\

\cline{1-1} \cline{2-5}

\\
\multicolumn{5}{c}{ \textit{$(a,\mu,\sigma,s)=(2,0,1,1)$} } \\
\\

$25 $ & $2.305\,(0.668)$ & $-0.0418\,(0.264)$ & $0.760\,(0.368)$  & $1.251\,(8.203)$ \\
$49 $ & $2.203\,(0.416)$ & $-0.0498\,(0.165)$ & $0.861\,(0.176)$  & $0.981\,(0.128)$ \\
$121$ & $2.097\,(0.160)$ & $-0.0318\,(0.063)$ & $0.944\,(0.066)$  & $0.991\,(0.024)$ \\
\\
\multicolumn{5}{c}{ \textit{$(a,\mu,\sigma,s)=(2,0,1,2)$}} \\
\\

$25 $ & $2.179\,(0.995)$ & $ 0.032\,(0.220)$  & $0.886\,(0.164)$  & $6.294\,(121.926)$ \\
$49 $ & $2.248\,(1.146)$ & $-0.024\,(0.253)$  & $0.943\,(0.093)$  & $2.907\,( 13.092)$ \\
$121$ & $2.254\,(1.151)$ & $-0.049\,(0.267)$  & $0.967\,(0.044)$  & $2.119\,(  0.460)$ \\

\\
\multicolumn{5}{c}{ \textit{$(a,\mu,\sigma,s)=(2,0,1,3)$}}
\\

$25 $ & $1.818\,(0.706)$ & $0.143\,(0.167)$ & $0.943\,(0.064)$ & $10.673\,(211.943)$ \\
$49 $ & $1.921\,(0.639)$ & $0.074\,(0.147)$ & $0.988\,(0.036)$ & $ 5.232\,( 33.199)$ \\
$121$ & $2.022\,(0.499)$ & $0.007\,(0.111)$ & $1.011\,(0.019)$ & $ 3.416\,(  1.040)$ \\

\\
\multicolumn{5}{c}{ \textit{$(a,\mu,\sigma,s)=(3,0,1,1)$}}
\\
$25 $ & $3.402\,(0.675)$ & $ 0.037\,(0.290)$ & $0.742\,(0.474)$ & $1.151\,(5.995)$ \\
$49 $ & $3.241\,(0.426)$ & $-0.005\,(0.152)$ & $0.840\,(0.221)$ & $0.953\,(0.098)$ \\
$121$ & $3.148\,(0.274)$ & $-0.010\,(0.078)$ & $0.901\,(0.097)$ & $0.966\,(0.021)$ \\

\\
\multicolumn{5}{c}{ \textit{$(a,\mu,\sigma,s)=(3,0,1,2)$}}
\\
$25 $ & $2.886\,(1.279)$ & $0.224\,(0.261)$ & $0.824\,(0.164)$ & $4.522\,(62.262)$ \\
$49 $ & $2.956\,(1.246)$ & $0.142\,(0.236)$ & $0.903\,(0.093)$ & $2.573\,( 7.336)$ \\
$121$ & $3.002\,(1.157)$ & $0.073\,(0.219)$ & $0.958\,(0.043)$ & $2.096\,( 0.369)$ \\

\\
\multicolumn{5}{c}{  \textit{$(a,\mu,\sigma,s)=(3,0,1,3)$}}
\\

$25 $ & $2.471\,(1.401)$  & $0.303\,(0.256)$  & $0.833\,(0.124)$  & $7.228\,(93.527)$  \\
$49 $ & $2.553\,(1.177)$  & $0.228\,(0.210)$  & $0.904\,(0.067)$  & $4.235\,(16.030)$  \\
$121$ & $2.710\,(0.921)$  & $0.134\,(0.157)$  & $0.957\,(0.037)$  & $3.228\,( 0.564)$  \\

\\ \bottomrule

\end{tabular}

\end{table*}

\subsection{SAR Image Modeling}

Polarimetric SAR (PolSAR) devices
have been considered
one of most important tools
for sensing geophysical scenarios~{\cite{LeePottier2009PolarimetricRadarImaging}}.
Such systems work
under
the following dynamic:
orthogonally polarized pulses
(in vertical, `V', or horizontal, `H', directions)
are transmitted
towards a target,
and the returned echoes are recorded
with respect to each polarization.
Resulting images
can be understood
as outcomes from
a sequence
either
(i)~of complex random vectors
(called \textit{single look})
or
(ii)~of Hermitian positive definite random matrices
(called \textit{multilook})~\cite{Akinseteetal}.
Here, we
address the multilook case,
considering
the proposal of statistical models
for elements of the main diagonal of
resulting matrices from
PolSAR images.

\begin{figure}[htbp]
\centering
\subfigure[Total image\label{Bias1}]{\includegraphics[width=.58\linewidth,height=.66\linewidth]{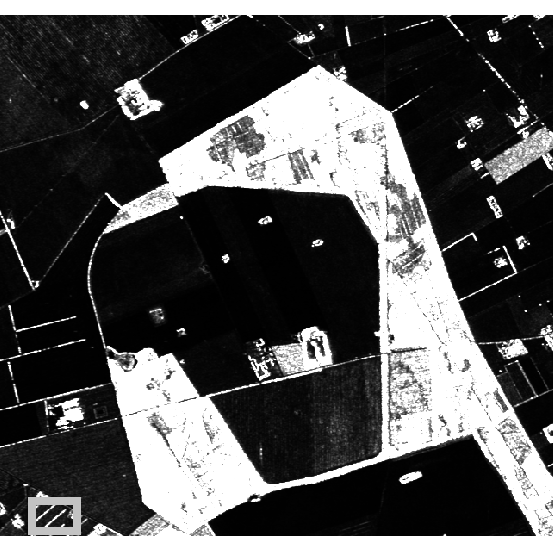}}
\subfigure[HV, VV, and HH region images\label{Bias22}]{\includegraphics[width=.38\linewidth]{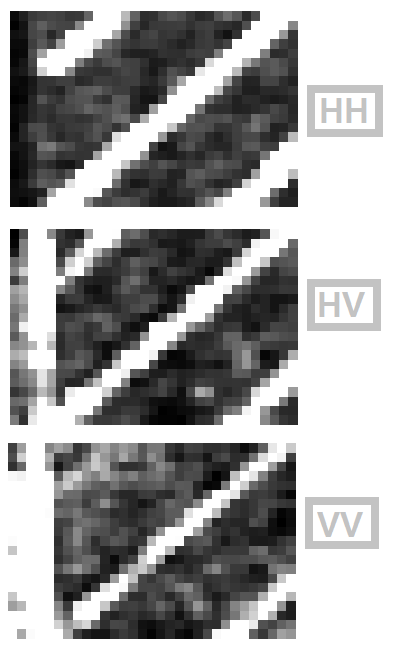}}\\
\caption{SAR image with the selected region and images of data of HH, HV, and VV associated polarization channels.}
\label{mainimage}
\end{figure}

Figure~\ref{Bias1}
presents
an image over the surroundings of
Foulum (Denmark) obtained by the EMISAR system~\cite{Doulgerisetal2011}.
This
is
a polarimetric SAR image,
i.e.,
each of its pixels
is represented
by
a particular
3$\times$3 Hermitian positive definite matrices,
whose diagonal elements are positive real intensities:
\[
\bm{Z}=
\begin{bmatrix}
|Z_\text{HH}| & Z_\text{HH-HV}  & Z_\text{HH-VV} \\
Z_\text{HH-HV}^* & |Z_\text{HV}|  & Z_\text{HV-VV} \\
Z_\text{HH-VV}^* & Z_\text{HV-VV}^*  & |Z_\text{VV}|
\end{bmatrix}
,
\]
where $\{|Z_\text{HH}|,$ $\,|Z_\text{HV}|,$ $\,|Z_\text{VV}|\}$ represents the set of intensities from $Z_\text{HH}$, $Z_\text{HV}$, and $Z_\text{VV}$ po\-la\-rization channels (complex random vectors)  and
$\{Z_\text{HH-HV},$ $\,Z_\text{HH-VV},$ $\,Z_\text{HV-VV}\}$ indicates the set of possible products between two different polarization channels such that $Z_\text{A-B}=Z_\text{A}Z_\text{B}^*$, for A,B $\in\{\text{HH}, \text{HV}, \text{VV}\}$,
and $^{*}$ denotes the conjugate of a complex number.
The intensities of the echoed
signal polarization channels play an important role,
since they depend on the physical properties of
the target surface.
We aim to show that the proposed model
is an adequate representation
for
the
a pre-processing step
in
SAR image processing due to one channel.
Figure~\ref{Bias22}
illustrates
the images of the corresponding polarization channels.

\begin{figure}
\centering

\subfigure[HH channel~\label{APLII:1:1}]{\includegraphics[width=.48\linewidth]{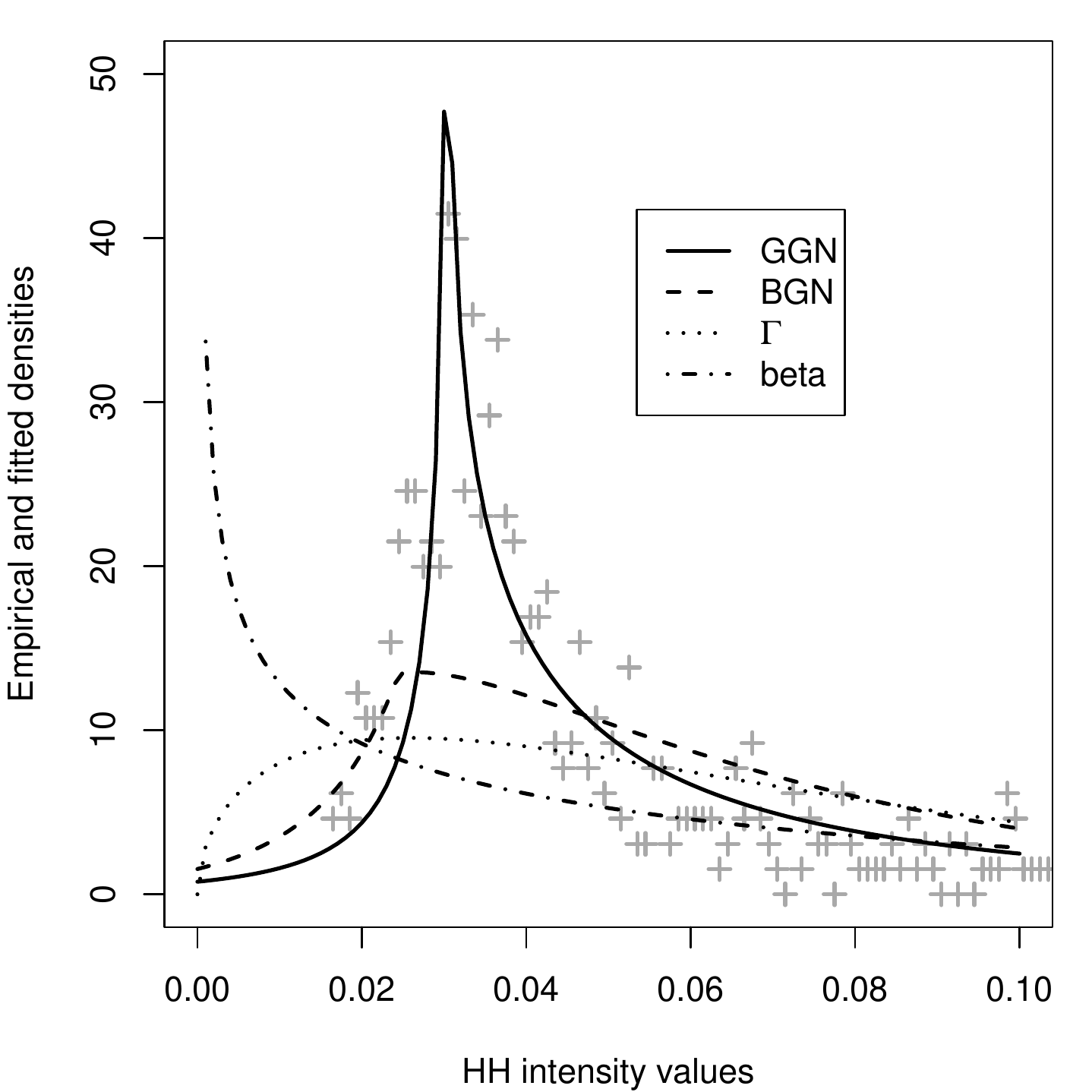}}
\subfigure[HV channel~\label{APLII:1:2}]{\includegraphics[width=.48\linewidth]{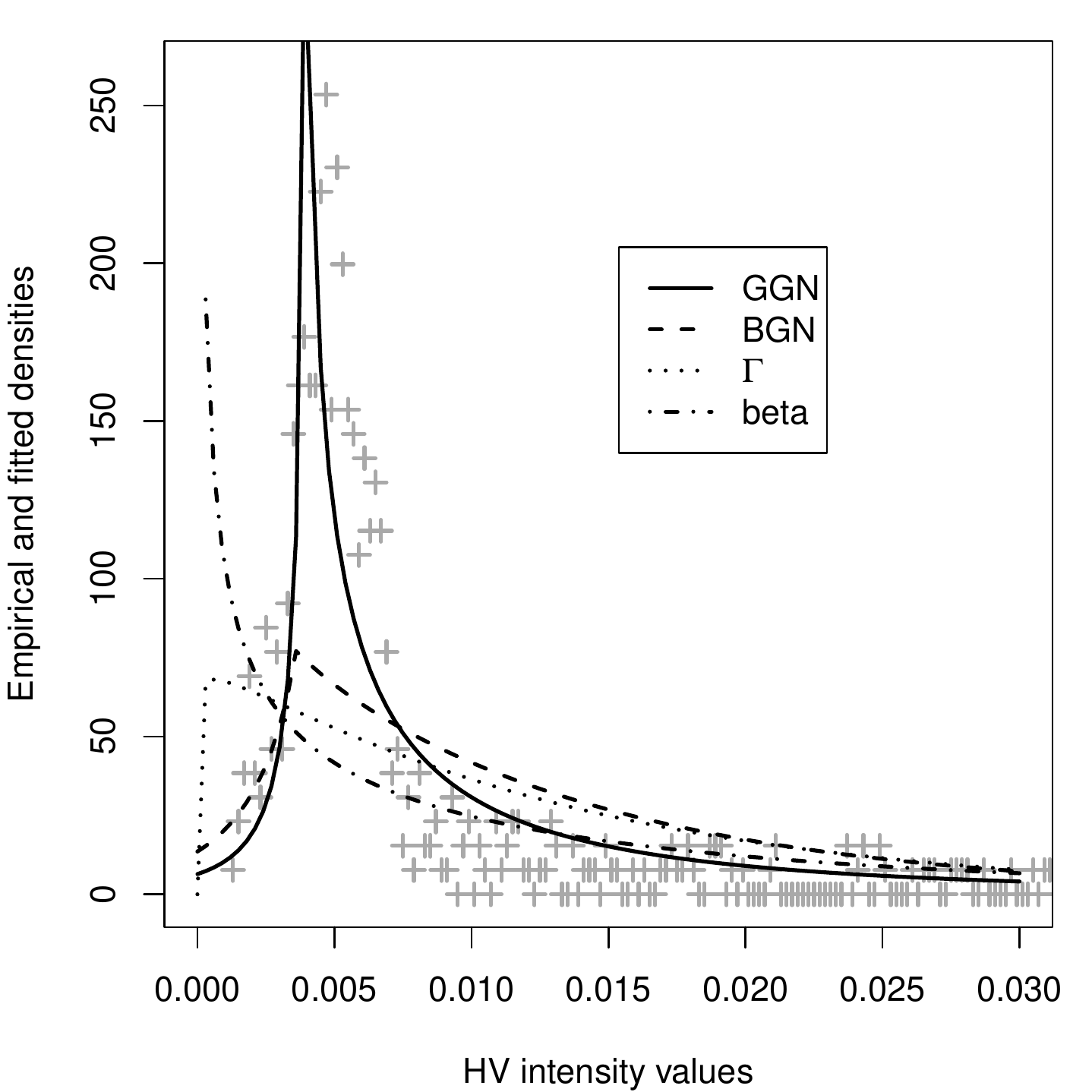}}
\subfigure[VV channel~\label{APLII:1:3}]{\includegraphics[width=.48\linewidth]{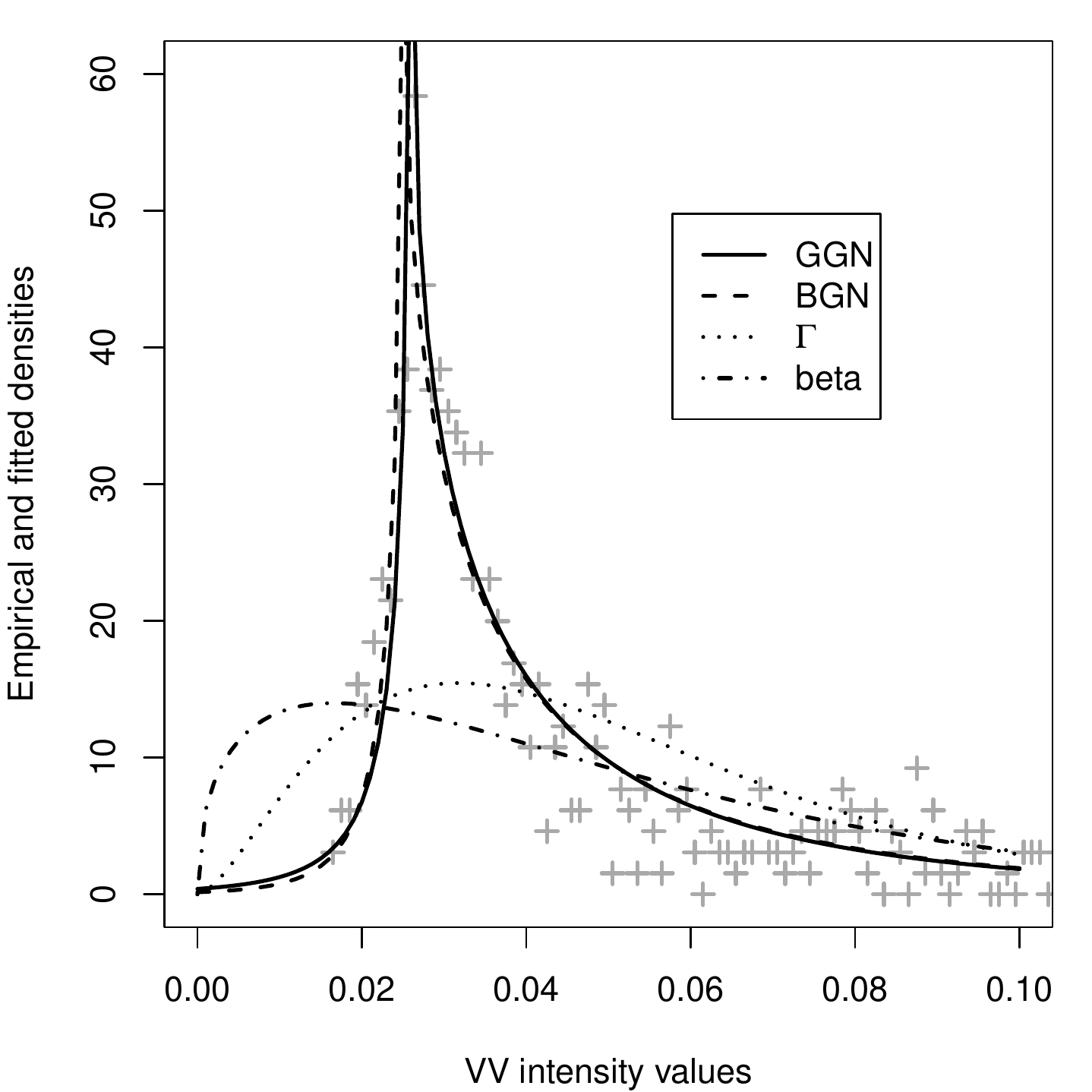}}

\caption{%
Empirical densities of the channel data.}
\label{APLII:1}
\end{figure}

Figure~\ref{APLII:1}
gives
empirical
and
fitted densities on EMISAR real data.
The GGN model is compared
with the BGN distribution~\cite{cintraetal2013}.
Such distribution
outperforms
classical models for intensity SAR data, such as
$\mathcal G^0_I$, $\mathcal K$, and $\Gamma$ distributions;
however, it has five parameters
which usually offers
computational
difficulties,
because the MLEs are defined as the solution of a system with five nonlinear equations.
{%
The fitted empirical density
plots shown in Figures~\ref{APLII:1:1}-\ref{APLII:1:3}
suggest the appropriateness of
the GGN model.
In comparison with the BGN, $\Gamma$, and beta distributions,
the GGN model
offers
better fitted empirical densities.}
{%
Table~\ref{EST:1}
presents
the corresponding MLEs
and their respective estimates for the mean square errors
for the parameters of GGN, BGN, $\Gamma$, and beta distributions
on data from each polarization channel.
}

\begin{table}[!h]

\centering

\caption{ML estimates for the $\operatorname{BGN}(s,\mu,\sigma,\alpha,\beta)$~\cite{cintraetal2013}, $\operatorname{GGN}(a,\mu,\sigma,s)$, $\Gamma(\alpha,\beta)$ and $\operatorname{beta}(\alpha,\beta)$ distributions. Standard errors are in parenthesis.}
\label{EST:1}
\footnotesize
\begin{tabular}{c@{\quad}c@{\,\,}c@{\,\,}c@{\,\,}c@{\,\,}c@{}}
\toprule
Model &  \multicolumn{5}{c}{Estimated Parameters} \\

\hline
\multicolumn{6}{c}{\emph{For the HH channel}}\\

BGN    & 0.995   &  2.448 $\times 10^{-2}$   & 1.972 $\times 10^{-2}$  &  1.498   &  0.407  \\
       &(0.128)  & (0.312 $\times 10^{-2}$) & (0.429 $\times 10^{-2}$) & (0.585)  & (0.098) \\

GGN    &  1.803  &  3.031 $\times 10^{-2}$  &  8.232 $\times 10^{-4}$  &  0.356   & $\bullet$  \\
       & (0.192) & (1.950 $\times 10^{-3}$) & (8.834 $\times 10^{-5}$) & (5.707 $\times 10^{-3}$) & $\bullet$  \\

$\Gamma$  &  1.517     &  19.946    & $\bullet$ & $\bullet$ & $\bullet$  \\
          & (0.086)    & (1.399)    & $\bullet$ & $\bullet$ & $\bullet$  \\

beta   &  0.607  &  7.372  & $\bullet$ & $\bullet$ & $\bullet$  \\
       & (0.044) & (0.663) & $\bullet$ & $\bullet$ & $\bullet$  \\

\hline

\multicolumn{6}{c}{\emph{For the HV channel}}\\

BGN    &  0.989  &  3.603 $\times 10^{-3}$  &  2.595 $\times 10^{-3}$  &  0.949  &  0.232   \\
       & (0.122) & (0.644 $\times 10^{-3}$) & (0.318 $\times 10^{-3}$) & (0.571) & (0.212) \\

GGN    &  1.821  &  3.965 $\times 10^{-3}$  &  5.031 $\times 10^{-5}$  &  3.084 $\times 10^{-1}$  & $\bullet$  \\
       & (0.113) & (2.602 $\times 10^{-4}$) & (9.410 $\times 10^{-6}$) & (8.373 $\times 10^{-3}$) & $\bullet$  \\

$\Gamma$  &  1.045  &  79.817 & $\bullet$ & $\bullet$ & $\bullet$  \\
          & (0.034) & (1.605) & $\bullet$ & $\bullet$ & $\bullet$  \\

beta   &  0.528  &  39.835  & $\bullet$ & $\bullet$ & $\bullet$  \\
       & (0.017) & (1.638)  &  $\bullet$ & $\bullet$ & $\bullet$  \\

\hline

\multicolumn{6}{c}{\emph{For the VV channel}}\\

BGN   &   0.634  &  2.499 $\times 10^{-2}$  &  2.107 $\times 10^{-3}$   &  1.243  &  0.285  \\
      &  (0.099) & (0.212 $\times 10^{-2}$) & (1.429 $\times 10^{-3}$) & (0.881) & (0.120)\\

GGN   &  1.955                   &  2.592 $\times 10^{-2}$  &  1.369 $\times 10^{-3}$  &  0.433420   & $\bullet$  \\
      & (7.089 $\times 10^{-2}$) & (4.965 $\times 10^{-4}$) & (3.347 $\times 10^{-5}$) & (4.771 $\times 10^{-3}$) & $\bullet$  \\

$\Gamma$ &  2.67                    &   52.590 & $\bullet$ & $\bullet$ & $\bullet$  \\
         & (8.578 $\times 10^{-2}$) &  (1.546) & $\bullet$ & $\bullet$ & $\bullet$  \\

beta  &  1.436                   &   26.848  & $\bullet$ & $\bullet$ & $\bullet$  \\
      & (9.029 $\times 10^{-2}$) &   (2.027) & $\bullet$ & $\bullet$ & $\bullet$  \\
\bottomrule

\end{tabular}

\end{table}

\begin{table*}[htbp]

\centering

\footnotesize

\caption{
Goodness-of-fit measures for the fitted BGN, GGN, $\Gamma$ and beta distributions on EMISAR real data
}

\label{table:resul:3}
\begin{tabular}{cc cccc}

\toprule
\multirow{3}{*}{Goodness-of-fit measures} & \multirow{3}{*}{Channel} & \multicolumn{4}{c}{{\it Performance for different model}}\\
\cmidrule(lr{.25em}){3-6}
& &   GGN    & BGN & $\Gamma$ & beta \\
\midrule
$\text{d}_\text{KL}(X)$& HH &    37.957 &   104.992 &   126.852 &   121.491 \\
$\text{d}_{\chi^2}(X)$   &    &   107.403 & 223.218 $\times 10^{2}$ &  578.139 $\times 10$ &   701.5747 \\
$\text{d}_\text{KS}$   &    &     0.053 &     0.163 &     0.196 &     0.304 \\
$\text{W}^*$           &    &     0.452 &     5.396 &     7.392 &     8.116 \\
$\text{A}^*$           &    &     3.450 &    29.159 &    39.597 &    43.120 \\
$\text{AIC}$           &    & -2.495 $\times 10^{3}$ & -2.225 $\times 10^{3}$ & -2.108 $\times 10^{3}$ & -1.844 $\times 10^{3}$ \\
$\text{AIC}_c$         &    & -2.493 $\times 10^{3}$ & -2.223 $\times 10^{3}$ & -2.106 $\times 10^{3}$ & -1.842 $\times 10^{3}$ \\
$\text{BIC}$           &    & -2.490 $\times 10^{3}$ & -2.222 $\times 10^{3}$ & -2.099 $\times 10^{3}$ & -1.835 $\times 10^{3}$ \\
\hline
$\text{d}_\text{KL}(X)$& HV &  371.397   &  898.627             &   962.826             &  849.346  \\
$\text{d}_{\chi^2}(X)$   &    &   912.831  &  297.285 $\times 10$ &   285.695 $\times 10$ &  225.040 $\times 10$ \\
$\text{d}_\text{KS}$   &    &     0.117  &     0.302            &     0.275             &     0.276 \\
$\text{W}^*$           &    &     1.563  &     9.112            &    11.219             &    11.246 \\
$\text{A}^*$           &    &     9.002  &    47.860            &    58.162             &    58.241 \\
$\text{AIC}$           &    & -4.777 $\times 10^{3}$ & -4.428 $\times 10^{3}$ & -4.340 $\times 10^{3}$ & -4.177 $\times 10^{3}$ \\
$\text{AIC}_c$         &    & -4.775 $\times 10^{3}$ & -4.426 $\times 10^{3}$ & -4.338 $\times 10^{3}$ & -4.175 $\times 10^{3}$ \\
$\text{BIC}$           &    & -4.772 $\times 10^{3}$ & -4.425 $\times 10^{3}$ & -4.331 $\times 10^{3}$ & -4.168 $\times 10^{3}$ \\
\hline
$\text{d}_\text{KL}(X)$& VV &    78.428              &   98.590                &   272.411              &   247.027 \\
$\text{d}_{\chi^2}(X)$   &    &   211.659              &  261.400                &  978.570 $\times 10$   &  102.669 $\times 10$ \\
$\text{d}_\text{KS}$   &    &     0.049              &  5.125 $\times 10^{-2}$ &     0.174              &     0.216 \\
$\text{W}^*$           &    &     0.262              &  0.247                  &     5.945              &     6.238 \\
$\text{A}^*$           &    &     1.900              &  1.514                  &    32.422              &    33.924 \\
$\text{AIC}$           &    & -3.191 $\times 10^{3}$ & -3.200 $\times 10^{3}$  & -2.847 $\times 10^{3}$ & -2.728 $\times 10^{3}$ \\
$\text{AIC}_c$         &    & -3.189 $\times 10^{3}$ & -3.198 $\times 10^{3}$  & -2.845 $\times 10^{3}$ & -2.726 $\times 10^{3}$ \\
$\text{BIC}$           &    & -3.186 $\times 10^{3}$ & -3.197 $\times 10^{3}$  & -2.838 $\times 10^{3}$ & -2.719 $\times 10^{3}$ \\
\bottomrule
\end{tabular}
\end{table*}

In order to compare the BGN and GGN models for the current data,
we consider two discrimination measures based on information-theoretical tools~\cite{est3,Taneja2006}.
Let $X$ be a random variable with density $f_X(x)$
and
$P_n$ the empirical density obtained from a data set.
The following sample goodness-of-fits measures
are considered:
\begin{itemize}
\item[a)] Symmetrized Kullback-Leibler divergence~\cite{est3,SeghouaneAmari2007}:
\begin{align*}
d_\text{KL}(X)\equiv &\, d_\text{KL}(X,P_n)\\
&=\,\sum_{i=1}^{N}\,\Bigg\{\,f_X(X_i;\widehat{\bm{\theta}}_x)\,\log\left[\frac{f_X(X_i;\widehat{\bm{\theta}}_x)}{P_i}\right]
\\
&\quad\,
+\,
\,{P_i}\,\log\left[\frac{P_i}{f_X(X_i;\widehat{\bm{\theta}}_x)}\right]\Bigg\}
\end{align*}
\item[b)] Symmetrized chi-square divergence~\cite{Taneja2006}:
\begin{align*}
d_{\chi^2}(X)\equiv& d_{\chi^2}(X,P_n)\,=\,\sum_{i=1}^{N}\,\Bigg[\,
\frac{(f_X(X_i;\widehat{\bm{\theta}}_x)-P_i)^2}{f_X(X_i;\widehat{\bm{\theta}}_x)}\\
&\,+\,
\frac{(f_X(X_i;\widehat{\bm{\theta}}_x)-P_i)^2}{P_i}
\Bigg].
\end{align*}

\end{itemize}

Additionally, we also use the
classical Kolmogorov--Smirnov ($\text{d}_\text{KS}$),
Anderson--Darling ($\text{A}^*$)
and
Cr\'amer--von~Mises ($\text{W}^*$)
statistics~\cite{Evans2008}.
These measures are more adequate
than
the Akaike information criterion (AIC),
corrected AIC (AIC$_c$),
and Bayesian information criterion (BIC) measures.
The latter ones are more recommended for nested models~\cite{Quang1989}.
Table~\ref{table:resul:3}
presents
the values
of these eight measures
for comparing
the GGN and BGN models
in terms of their
distributions from intensity SAR data.

{%
In general,
comparisons
favor the GGN model.
An exception to this behavior
is
the group of measures
$\{\text{AIC}, \text{AIC}_c, \text{BIC}\}$
for the VV channel.
In this case,
the BGN distribution
is
the best descriptor of SAR VV channel
intensities.
However,
according to
Voung~\cite{Quang1989},
the group $\{\text{AIC}, \text{AIC}_c, \text{BIC}\}$
is more suitable for nested models,
which is not the case
in the GGN, BGN, $\Gamma$, and beta models.
As a consequence,
we suggest
the GGN model
as a meaningful model
to describe SAR data from HH and VV channels.
}

\subsection{AIRSAR Sensor Imagery Analysis}

\begin{figure}
\centering

\subfigure[AIRSAR image~\label{APLII:2:1}]{\includegraphics[width=.8\linewidth]{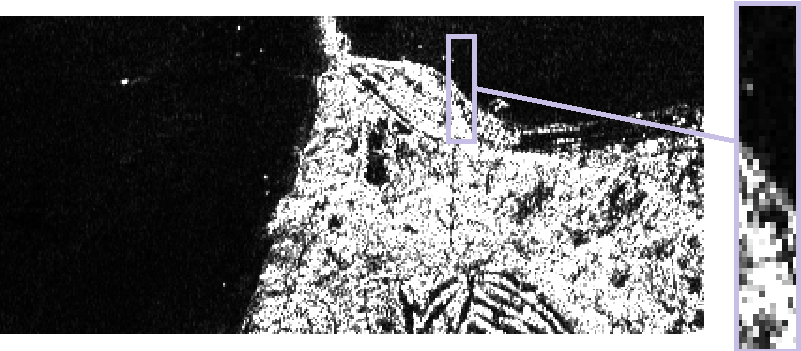}}\\
\subfigure[HH channel~\label{APLII:2:2}]{\includegraphics[width=.38\linewidth]{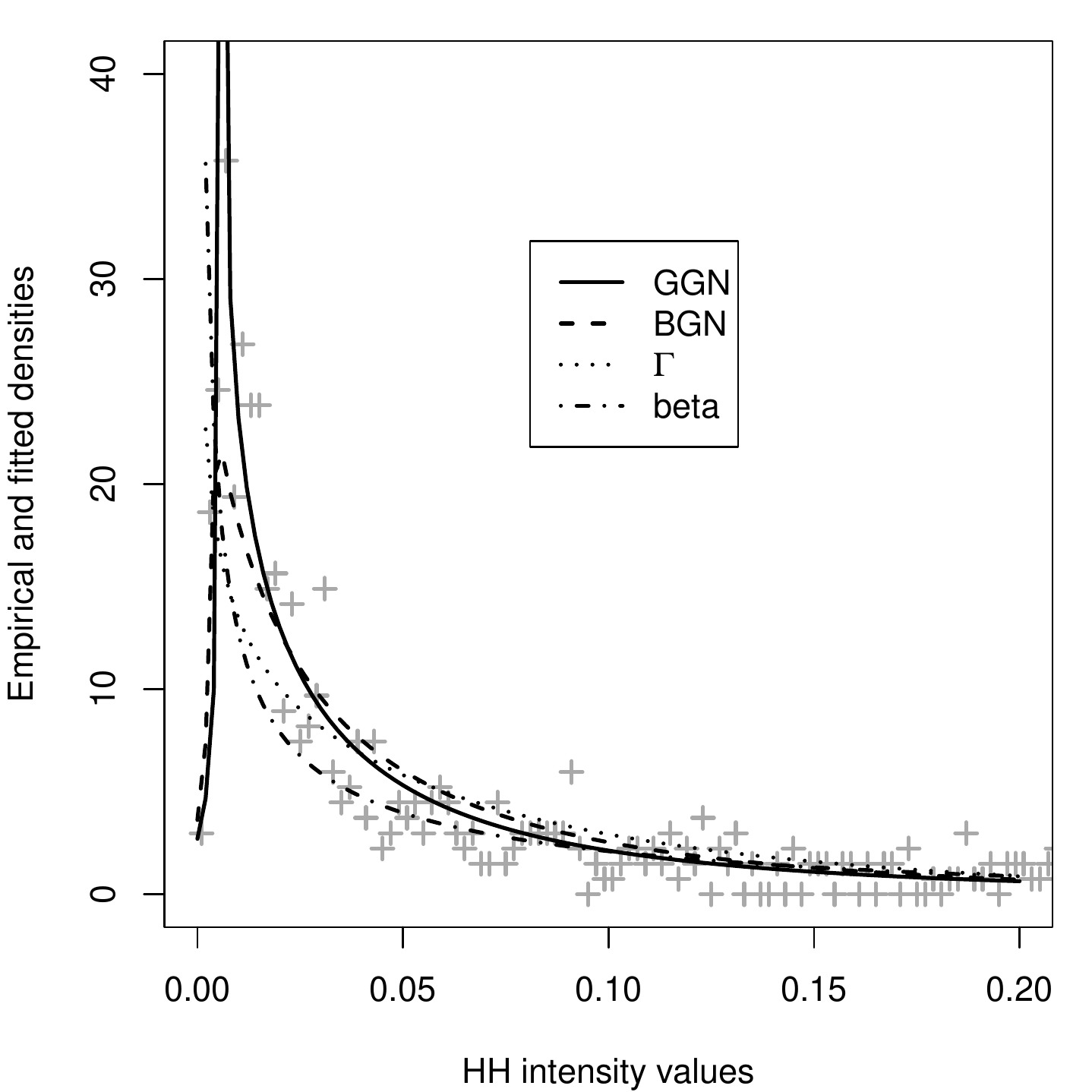}}
\subfigure[HV channel~\label{APLII:2:3}]{\includegraphics[width=.38\linewidth]{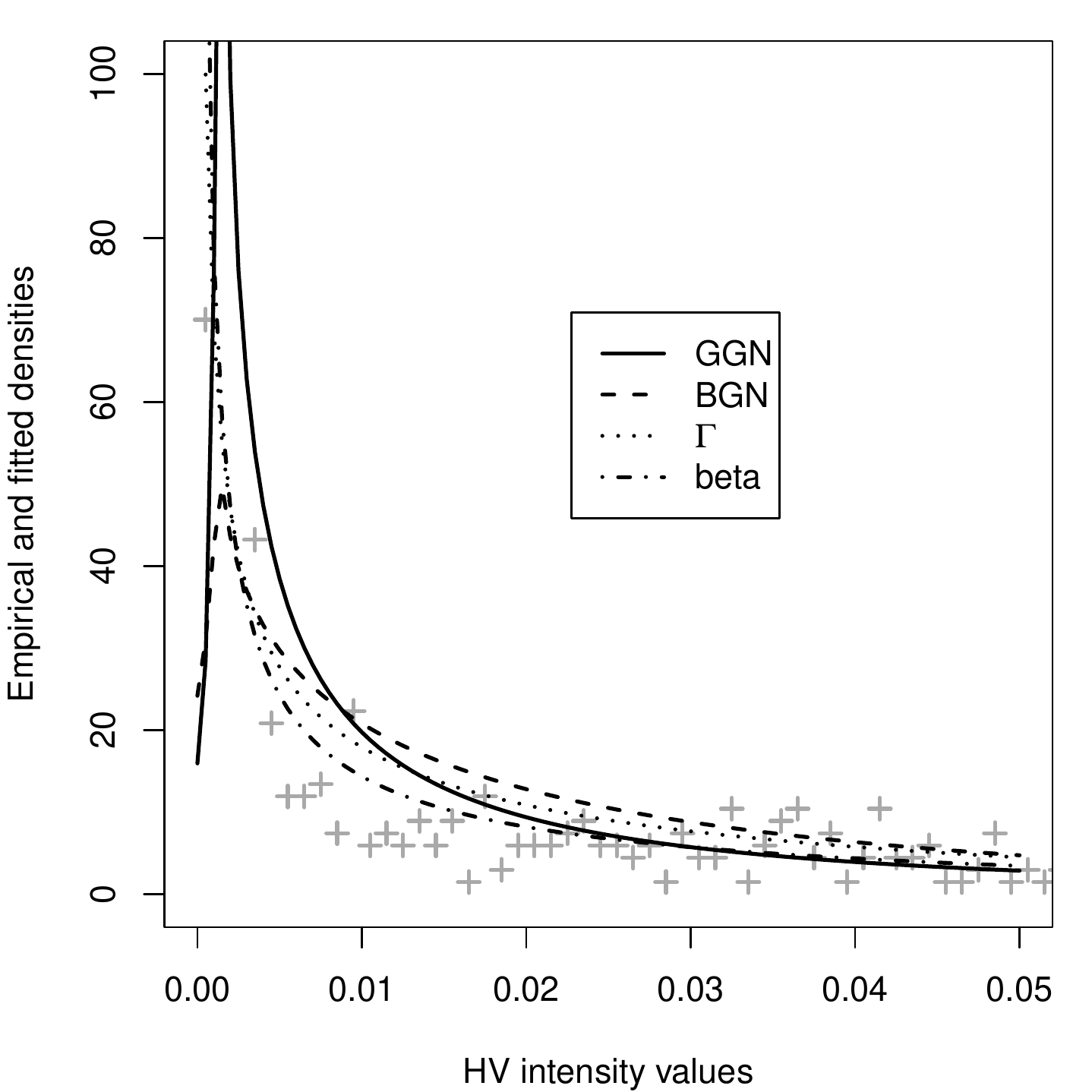}}
\subfigure[VV channel~\label{APLII:2:4}]{\includegraphics[width=.38\linewidth]{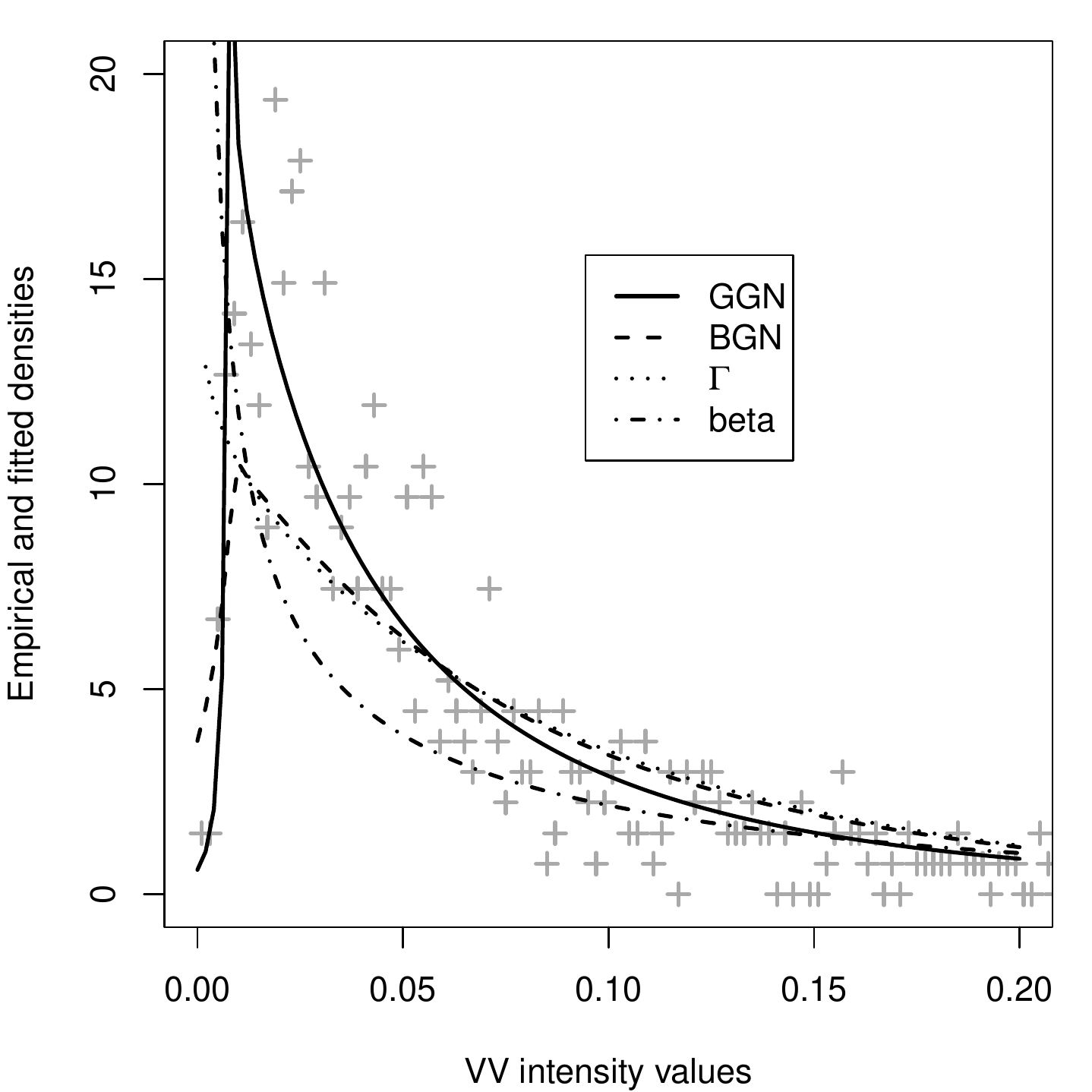}}

\caption{%
(a)~AIRSAR image with a selected strip
and
(b)--(c)~fitted and empirical densities
for the HH, HV, and VV channels,
respectively.}
\label{APLII:2}
\end{figure}

In this section,
we apply the proposed methodology
to a San Francisco Bay image
acquired
from an AIRSAR sensor with the number of looks equal to four~\cite{Nascimento2014}.
Fig.~\ref{APLII:2:1} shows the adopted SAR image for the HV channel
and a particular selected image strip for analysis.
Now we aim at describing
the intensities of the imagery within the selected
strip according to two regions:
forest and ocean.
Fitted data shown in Fig.~\ref{APLII:2:2}--\ref{APLII:2:4}
display
a qualitative comparison
favorable to
the BGN and GGN models,
which a capable of better fitting than
$\Gamma$ and beta standard distributions.
In particular,
the proposed GGN
seems to outperform the competing models.
In a quantitative analysis,
ML estimates involved in this application
are presented in
Table~\ref{EST:2}.
Table~\ref{comparison:APLIII}
shows goodness-of-fit values.
The GGN distribution
is
the best model
according to all adopted criteria.

\begin{table}[!h]
\centering
\caption{ML estimates for the $\operatorname{BGN}(s,\mu,\sigma,\alpha,\beta)$~\cite{cintraetal2013}, $\operatorname{GGN}(a,\mu,\sigma,s)$, $\Gamma(\alpha,\beta)$ and $\operatorname{beta}(\alpha,\beta)$ distributions. Standard errors are in parenthesis. }
\label{EST:2}
\footnotesize
\begin{tabular}{c@{\quad}c@{\,\,}c@{\,\,}c@{\,\,}c@{\,\,}c@{}}
\toprule
Model &  \multicolumn{5}{c}{Estimated Parmeters} \\
\hline
\multicolumn{6}{c}{\emph{For the HH channel}}\\

BGN    &  0.665  & 4.291 $\times 10^{-2}$  &  3.240 $\times 10^{-2}$  &  1.675   &  0.188 \\
       & (0.101) &(1.655 $\times 10^{-2}$) & (3.336 $\times 10^{-2}$) & (0.821)  & (0.116)\\

GGN    & 2.791   &  4.283 $\times 10^{-2}$   & 1.496 $\times 10^{-3}$   &  0.304   & $\bullet$  \\
       &(0.175)  & (1.151 $\times 10^{-2}$) & (1.661 $\times 10^{-3}$)  & (0.016)  & $\bullet$  \\

$\Gamma$  & 0.725  &   9.935  & $\bullet$ & $\bullet$ & $\bullet$  \\
         & (0.034) &  (0.637) & $\bullet$ & $\bullet$ & $\bullet$  \\

beta   &  0.376  &  4.775  & $\bullet$ & $\bullet$ & $\bullet$  \\
       & (0.049) & (1.041) & $\bullet$ & $\bullet$ & $\bullet$  \\
\hline
\multicolumn{6}{c}{\emph{For the HV channel}}\\

BGN    &  0.713  &  1.283  $\times 10^{-2}$   &  2.362 $\times 10^{-3}$  &  0.962  &  0.186 \\
       & (0.191) & (0.1693 $\times 10^{-2}$)  & (2.741 $\times 10^{-3}$) & (0.292) & (0.066)\\

GGN    &  2.313  &  1.198 $\times 10^{-3}$  &  2.321 $\times 10^{-6}$  & 2.157  $\times 10^{-1}$ & $\bullet$  \\
       & (0.128) & (1.647 $\times 10^{-4}$) & (7.479 $\times 10^{-6}$) & (1.595 $\times 10^{-2}$) & $\bullet$  \\

$\Gamma$  &  0.468   & 13.152  & $\bullet$ & $\bullet$ & $\bullet$  \\
          & (0.022)  & (1.063) & $\bullet$ & $\bullet$ & $\bullet$  \\

beta   &  0.296   &  8.020  & $\bullet$ & $\bullet$ & $\bullet$  \\
       & (0.021)  & (0.904) &  $\bullet$ & $\bullet$ & $\bullet$  \\
\hline
\multicolumn{6}{c}{\emph{For the VV channel}}\\

BGN   &  0.857  &  9.104 $\times 10^{-3}$  &  1.552 $\times 10^{-2}$  & 1.369    &  0.251   \\
      & (0.107) & (2.915 $\times 10^{-3}$) & (0.291 $\times 10^{-2}$) & (0.793)  & (0.120) \\

GGN   &  2.766  &  8.894 $\times 10^{-2}$  &  0.907 $\times 10^{-3}$ &  0.373  & $\bullet$  \\
      & (0.403) & (4.375 $\times 10^{-2}$) & (1.002 $\times 10^{-3}$)& (0.029) & $\bullet$  \\

$\Gamma$ & 0.388  &  3.929  & $\bullet$ & $\bullet$ & $\bullet$  \\
      &   (0.044) & (0.636) & $\bullet$ & $\bullet$ & $\bullet$  \\

beta  &  0.925  &  10.280  & $\bullet$ & $\bullet$ & $\bullet$  \\
      & (0.071) &  (0.945) & $\bullet$ & $\bullet$ & $\bullet$  \\
\bottomrule
\end{tabular}
\end{table}

\begin{table*}[htbp]

\centering
\footnotesize

\caption{
Goodness-of-fit measures for the fitted BGN, GGN, $\Gamma$ and beta distributions on AIRSAR real data
}

\label{comparison:APLIII}

\begin{tabular}{cc cccc}

\toprule
\multirow{3}{*}{Goodness-of-fit measures} & \multirow{3}{*}{Channel} & \multicolumn{4}{c}{{\it Performance for different model}}\\
\cmidrule(lr{.25em}){3-6}
& &   GGN    & BGN & $\Gamma$ & beta \\
\midrule
$\text{d}_\text{KL}(X)$& HH &  32.440                 &  37.960                 &    56.361              &    59.518              \\
$\text{d}_{\chi^2}(X)$   &    &  84.720                 &  112.800                &   211.733              &   167.780              \\
$\text{d}_\text{KS}$   &    &  4.501 $\times 10^{-2}$ &  9.162 $\times 10^{-2}$ &     0.125              &     0.206              \\
$\text{W}^*$           &    &  0.312                  &  0.791                  &     2.307              &     2.610              \\
$\text{A}^*$           &    &  1.772                  &  4.506                  &    13.458              &    15.189               \\
$\text{AIC}$           &    & -2.332 $\times 10^{3}$  & -2.309 $\times 10^{3}$  & -2.218 $\times 10^{3}$ & -2.055 $\times 10^{3}$  \\
$\text{AIC}_c$         &    & -2.329 $\times 10^{3}$  & -2.307 $\times 10^{3}$  & -2.216 $\times 10^{3}$ & -2.053 $\times 10^{3}$  \\
$\text{BIC}$           &    & -2.327 $\times 10^{3}$  & -2.306 $\times 10^{3}$  & -2.209 $\times 10^{3}$ & -2.046 $\times 10^{3}$  \\
\hline
$\text{d}_\text{KL}(X)$& HV &                 46.322  &    98.0581             &    79.318               &    67.2000              \\
$\text{d}_{\chi^2}(X)$   &    &                108.246  &   263.9445             &   203.313               &   158.1546              \\
$\text{d}_\text{KS}$   &    &                  0.115  &     0.2187             &     0.175               &     0.2153              \\
$\text{W}^*$           &    &                  1.823  &     4.4522             &     3.972               &     3.9070              \\
$\text{A}^*$           &    &                  9.738  &    26.4101             &    23.514               &    23.2618              \\
$\text{AIC}$           &    & -3.724 $\times 10^{3}$  & -3.375 $\times 10^{3}$ & -3.489 $\times 10^{3}$  & -3.399 $\times 10^{3}$  \\
$\text{AIC}_c$         &    & -3.722 $\times 10^{3}$  & -3.373 $\times 10^{3}$ & -3.487 $\times 10^{3}$  & -3.397 $\times 10^{3}$  \\
$\text{BIC}$           &    & -3.719 $\times 10^{3}$  & -3.372 $\times 10^{3}$ & -3.480 $\times 10^{3}$  & -3.390 $\times 10^{3}$  \\
\hline
$\text{d}_\text{KL}(X)$& VV &   14.880                &      21.124            &    26.523              &    36.672               \\
$\text{d}_{\chi^2}(X)$   &    &   40.420                &      88.446            &   223.184              &    86.066               \\
$\text{d}_\text{KS}$   &    &  4.468 $\times 10^{-2}$ &       0.118            &     0.120              &     0.243               \\
$\text{W}^*$           &    &   0.248                 &     1.791              &     2.243              &     2.907               \\
$\text{A}^*$           &    &   1.463                 &    10.911              &    13.294              &    17.050               \\
$\text{AIC}$           &    &  -2.009 $\times 10^{3}$ & -1.886 $\times 10^{3}$ & -1.889 $\times 10^{3}$ & -1.641 $\times 10^{3}$  \\
$\text{AIC}_c$         &    &  -2.007 $\times 10^{3}$ & -1.884 $\times 10^{3}$ & -1.887 $\times 10^{3}$ & -1.639 $\times 10^{3}$  \\
$\text{BIC}$           &    &  -2.004 $\times 10^{3}$ & -1.883 $\times 10^{3}$ & -1.880 $\times 10^{3}$ & -1.632 $\times 10^{3}$  \\
\bottomrule
\end{tabular}
\end{table*}

Computational cost
is also quantified for both generalized models GGN and BGN.
Computations
were performed
in a 64-bit Intel(R) Core(TM) i5-3317U CPU
running at~1.70~GHz.
Table~\ref{TC:2}
summarizes
the
average execution times in seconds.
Such average times were obtained
from one hundred instantiations without replacement
with sample sizes of 100, 200, and 300
from selected AIRSAR region,
as suggested by Efron~\cite{Efron1990}.
As expected,
the execution times
increase
according to the sample size.
However,
the GGN estimation
could be computed faster
when compared to the BGN model computation.

\begin{table}
\centering
\caption{Computational time (in seconds) for the fitting according to the GGN and BGN models}
\label{TC:2}
\footnotesize
\begin{tabular}{cccc}
\toprule
Channel & Size  &    GGN  &    BGN \\
\hline
HH & 300 &   2.220   &   2.870  \\
   & 200 &   1.469  &   2.050  \\
   & 100 &   0.948 &   1.168 \\
\hline
HV & 300 &  2.150 &  2.709 \\
   & 200 &  1.754 &  2.303 \\
   & 100 &  1.108  &  1.239 \\
\hline
VV & 300 &  2.336  & 2.936 \\
   & 200 &  1.582 & 2.058 \\
   & 100 &  0.927 & 1.121 \\
\bottomrule
\end{tabular}
\end{table}

\section{Conclusion}
\label{GGN:9}

The new gamma generalized normal (GGN) distribution
is proposed and discussed.
Some statistical properties of the new model
are investigated, such as: asymptotic behavior,
density and moments, power series expansions, maximum likelihood estimation,
and random number generation.
The derived tools
bring themselves
a great potentiality
in both
future theoretical studies
and
applications
as input
in image post-processing activities:
\begin{itemize}
\item[i)]
Improvement in
the estimation of the shape parameter $s$
(which imposed the more hard estimation by our synthetic results)
of the proposed model
by using
the Mellin transform according ti Gao \emph{et al.}~\cite{Gao2018}.
\item[ii)]
Retrievement of images
via
the gamma generalized Gaussian distributions
equipped support vector machine,
analogously what was proposed by Ranjani and Babu~\cite{IJST108026}
and
\item[iii)]
Flexibilization of the change detection methods
through
the GNG model,
by modifying the proposal in~\cite{chavenew}.
\end{itemize}

Im\-por\-tantly,
evidence was found for applicability of the proposed GGN distribution in the modeling
of SAR images.
Real data analysis shows that the introduced model
can accurately characterize SAR intensities
for several polarization channels.
Moreover,
the GGN can outperform the beta generalized normal (BGN) distribution
in terms of goodness-of-fit statistics.
{%
Bayesian restoration methods for SAR images
(such as discussed in~\cite{BoumanSauer1993})
involving the GGN and BGN models
is
a future research line
along with
other post-processing steps.
}

\section*{Acknowledgements}
The authors would like to thank the financial support of CNPq and FACEPE, Brazil.
Additionally, we also would like to thank to the European Space Agency (ESA), which has distributed software and
sources of datasets in form of the Polarimetric SAR Data Processing and Educational (PolSARpro) tool.
%

%

%

% \onecolumn

{\small
\singlespacing
\bibliographystyle{siam}
\bibliography{GGN}
}

\end{document}